\documentclass[runningheads]{llncs}
\usepackage[T1]{fontenc}
\usepackage{graphicx}
\usepackage{amsmath,amsfonts}
\usepackage{xcolor}

\usepackage{hyperref}
\hypersetup{
  colorlinks,
  citecolor=blue,
  linkcolor=red,
  urlcolor=blue
}

\usepackage{enumitem}
\setlist{nolistsep}

\begin{document}
%
\title{MetaRegNet: Metamorphic Image Registration Using Flow-Driven Residual Networks}
\titlerunning{MetaRegNet}
%
\author{Ankita Joshi\inst{1}, Yi Hong\inst{2}}
\authorrunning{Joshi and Hong}
%
\institute{Department of Computer Science, University of Georgia, Athens, GA, 30602, USA \and Department of Computer Science and Engineering, Shanghai Jiao Tong University, Shanghai, 200240, China\\
\email{yi.hong@sjtu.edu.cn}}
\maketitle              
\begin{abstract}
Deep learning based methods provide efficient solutions to medical image registration, including the challenging problem of diffeomorphic image registration. However, most methods register normal image pairs, facing difficulty handling those with missing correspondences, e.g., in the presence of pathology like tumors. We desire an efficient solution to jointly account for spatial deformations and appearance changes in the pathological regions where the correspondences are missing, i.e., finding a solution to metamorphic image registration. Some approaches are proposed to tackle this problem, but they cannot properly handle large pathological regions and deformations around pathologies. In this paper, we propose a deep metamorphic image registration network (MetaRegNet), which adopts time-varying flows to drive spatial diffeomorphic deformations and generate intensity variations. We evaluate MetaRegNet on two datasets, i.e., BraTS 2021 with brain tumors and 3D-IRCADb-01 with liver tumors, showing promising results in registering a healthy and tumor image pair. The source code is available online. 
\end{abstract}
\section{Introduction}

Deformable image registration (DIR) establishes pixel/voxel dense correspondences between 2D/3D images using a deformation that transforms images into a common space for fusion or comparison~\cite{sotiras2013deformable}. Existing DIR methods include classical registration models, e.g. SyN~\cite{avants2011reproducible}, Large Deformation Diffeomorphic Metric Mapping (LDDMM)~\cite{beg2005computing,cao2005large}, Stationary Velocity Fields (SVF)~\cite{arsigny2006log}, and recent deep-learning-based methods, e.g., QuickSilver~\cite{yang2017quicksilver}, VoxelMorph~\cite{dalca2018unsupervised}, SYMNet~\cite{mok2020fast}. These methods focus on registering image pairs with no missing correspondences, i.e., all pixels or voxels in the source image are matched with those in the target image using a bijective mapping function. Such assumptions limit their ability to tackle registration between image pairs with appearing or disappearing structures, e.g., developing brain scans during myelination, a healthy image and a tumor one, etc. These image pairs have both spatial deformation caused by the movements of shared structures and intensity changes caused by missing correspondences between unshared ones. Such intensity changes challenge existing methods, including diffeomorphic image registration that provides smooth transformation with a smooth inverse to preserve topology between image pairs.   

To capture spatial and intensity changes simultaneously, we turn to metamorphic image registration~\cite{trouve2005metamorphoses}, which introduces a source term to simulate the intensity changes in the diffeomorphic image registration framework~\cite{hong2012metamorphic}. The model complexity of traditional metamorphic image registration makes them impractical when handling large-scale and high-resolution images. To reduce the computational cost of existing methods, researchers propose some deep-learning-based solutions to handle pathological image registration, a special case of deep metamorphic image registration, e.g., the existence of tumor regions as in~\cite{maillard2022deep,mok2022unsupervised}. Existing methods use either a cost function masking (CFM) strategy, which completely separates deformations and intensity changes in the pathological and non-pathological regions, or a clean/healthy source or target image where a tumor is simulated to match with the other one with tumor. These methods ignore the deformations of healthy regions effected by the pathological regions, which causes large artifacts within and surrounding the pathological regions. 

\begin{figure}[t]
\includegraphics[width=0.99\textwidth]{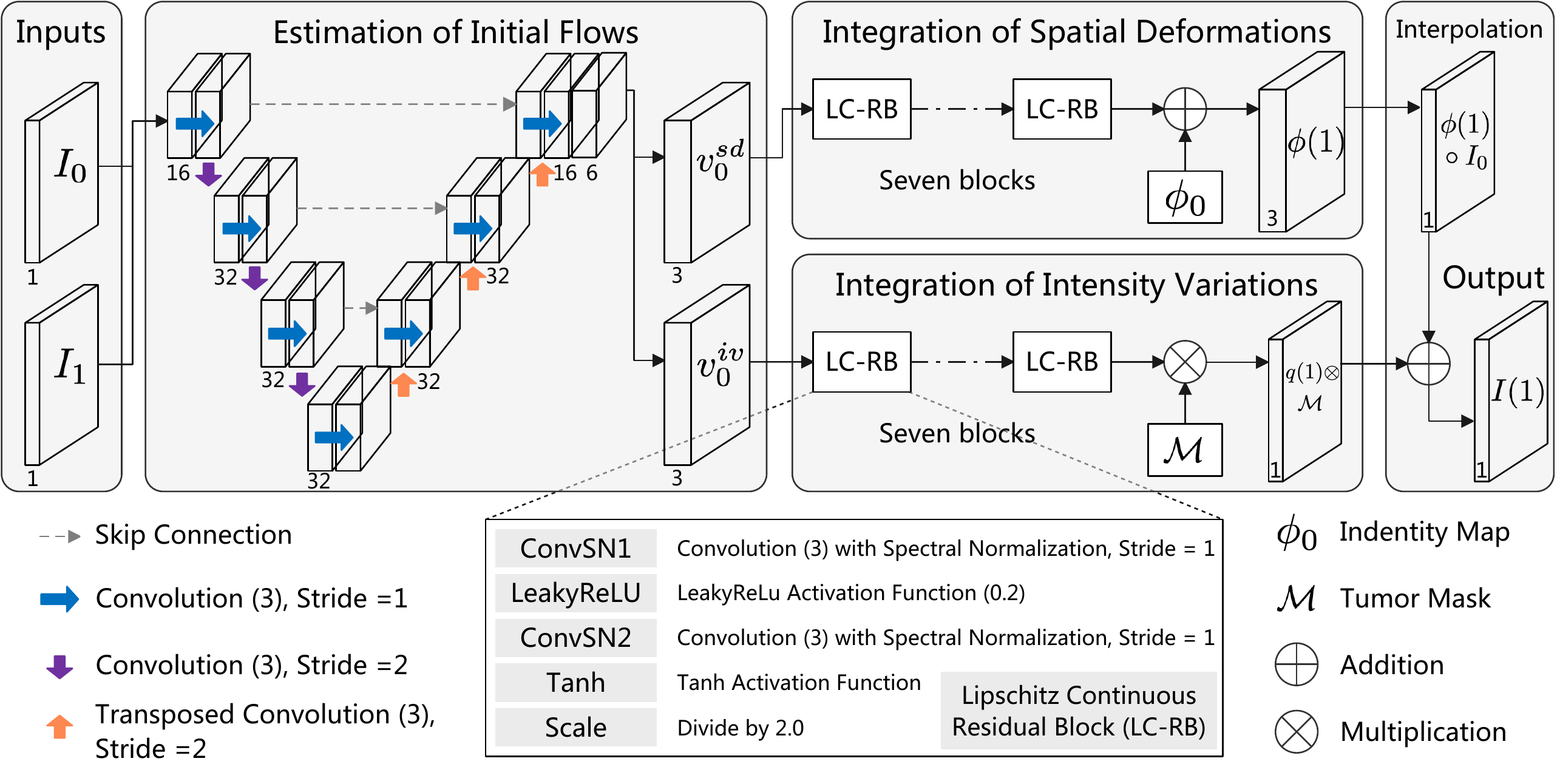}
\caption{Overview of our metamorphic image registration network, MetaRegNet.}
\label{fig:metaregnet}
\end{figure}

To address this challenge, we reformulate traditional metamorphic image registration and propose a deep metamorphic solution MetaRegNet, see Fig.~\ref{fig:metaregnet}. Based on Lipschitz continuous ResNet blocks proposed in~\cite{joshi2021diffeomorphic}, MetaRegNet consists of two pathways for jointly integrating spatial deformations and intensity changes along the trajectory from the source to target images. That is, MetaRegNet produces diffeomorphic deformations to account for spatial deformations between images, and jointly learns intensity changes to account for the tumor appearance between them. In this paper, we simplify the problem and limit our task to registering a healthy source to a pathological target, leaving other cases for future work. Our contributions in this paper are summarized below:
\begin{itemize}
    \item We propose a novel registration network for image-to-image metamorphosis, which uses the time-varying flow to drive diffeomorphic spatial mappings and simultaneously learns the incremental intensity variations between a healthy source image and a target image with one or more pathological regions.
    \item We conduct experiments on the BraTS 2021 brain tumor MRI~\cite{menze2014multimodal} and 3D-IRCADb-01 liver CT~\cite{3dircadb} datasets and compare to a metamorphic version of VoxelMorph~\cite{dalca2018unsupervised}. Our method produces significantly better results qualitatively and quantitatively, for both estimations of diffeomorphic mappings for spatial alignment and intensity variations in pathological regions. 
\end{itemize}

\subsection{Related Work}

\noindent
\textbf{Mask-based Methods.} 
This category assumes that pathological masks are available for learning; so that, the healthy and pathological regions can be treated separately. In~\cite{brett2001spatial}, masks are used to exclude the pathological regions from measuring the image similarity loss. The geometric metamorphosis~\cite{niethammer2011geometric} uses masks to separate the foreground and background deformation models for capturing intensity and spatial changes separately. 
Similarly, masking strategies have been used in~\cite{clatz2005robust,mok2022unsupervised}. Joint registration and segmentation approaches are also proposed in~\cite{chitphakdithai2010non,gooya2011joint,mok2022unsupervised}. These approaches assume that both source and target images have tumors and their models are hard to optimize and need a long time to converge.

Differently, we automatically learns the amount of intensity changes and how to balance them with spatial deformations. The mask {\it softly} restricts the learning of intensity changes within the pathological regions, resulting continuous deformations and greatly reduced artifacts surrounding the pathological regions. 



\noindent
\textbf{Reconstruction-based Methods.} Another choice for metamorphic image registration is to reconstruct a pathological image into a clean quasi-normal one before registration~\cite{han2018brain,bone2020learning,han2020deep}. A variety of techniques have been proposed, such as image in-painting~\cite{sdika2009nonrigid}, Variational AutoEncoder (VAE) based approaches~\cite{bone2020learning}, and a low-rank decomposition model to learn a normal image appearance~\cite{liu2014low}. These approaches do not require masks of pathological regions; however, their registration quality is limited by the imperfect or over-smoothed reconstructions. Also, these methods require extra healthy image scans for training and work well only if the size of the pathological region, like tumor or lesion, is relatively small. 


\noindent
\textbf{Other Methods}. In~\cite{bone2020learning,shu2018deforming,maillard2022deep}, researchers disentangle the shape and appearance changes of an image when performing registration.  In~\cite{gooya2012glistr,franccois2022weighted}, a biophysical model is adopted to introduce a growing tumor into a healthy image and perform image registration with another tumor image; however, modeling the growth of a tumor is a non-trivial task. In \cite{franccois2021metamorphic}, a semi-langrangian scheme is proposed to carry out registration for each pair but has limited ability to handle a large-scale dataset.

The model in~\cite{maillard2022deep} is the closest one to ours. It also uses residual networks and deforms a pathological source image to a fixed healthy atlas, which is a special case of ours, since we allow choosing different healthy images. More importantly, unlike them, we do not need a hyper-parameter to balance spatial deformations and intensity changes, which is not practical since it varies with different inputs.

\section{Background and Reformulation}
\label{sec:background}
In the LDDMM framework~\cite{beg2005computing}, diffeomorphic image registration is formulated as a minimization problem, which estimates a smooth deformation field $\phi: \Omega \rightarrow \Omega$, where $\Omega \subseteq R^ {n_{x} \times n_{y} \times n_{z}}$ (i.e., the size of a 3D image). The formulation is given as
\begin{equation}
    E(v) = \frac{1}{2}  \int_{0}^{1} {\| v \|_{L}^{2}}\,dt + \frac{1}{\sigma^{2}} {\|I_{1} - I(1) \|}_2^2, \quad s.t. \ I_{t} + \nabla I^Tv = 0, I(0)= I_{0}.
\label{eq:lddmm}    
\end{equation}
Here, $v$ is the time-dependent velocity field, $L$ is a spatial regularizer to ensure the smoothness of $v$; $I_0$ and $I_1$ are the source and target images, $I(0)$ and $I(1)$ are the deformed images at $t=0$ and $t=1$, respectively; $\sigma$ controls the influence of the regularization term and the image matching term. This formulation is an image-based version of LDDMM, where the image intensity is driven by the velocity field $v$. We can also use map-based implementation, which uses a deformation $\phi$ that is driven by the velocity field and then used to warp the source image. 

Metamorphic registration can be formulated based on the diffeomorphic LDDMM model~\cite{hong2012metamorphic}, by introducing a control variable $q$ in the image transport equation, as shown in Eq.~\eqref{eq:metamorphosis}. This introduced variable $q$ simulates an intensity source term and models intensity changes caused by appearing or disappearing objects in the image. The new optimization function is formulated as 
\begin{equation}
    E(v,q) = \frac{1}{2}\int_{0}^{1} {\| v \|_{L}^{2}}\,dt + \rho{\| q \|_{Q}^{2}}\,dt \ s.t. \ I_{t} + \nabla I^Tv = q, I(0)= I_{0}, I(1) = I_{1},
\label{eq:metamorphosis}    
\end{equation}
where $\rho$ is a constant value to balance the intensity variations introduced by the control variable $q$, $Q$ is a smooth operator applied on $q$. Different from LDDMM, metamorphic registration moves the image matching term into the constraints and can achieve a {\it perfect matching} between warped image $I(1)$ and the target image $I_1$, due to the introduced $q$. The solution in~\cite{hong2012metamorphic} presents a tight coupling of velocity fields that drive spatial deformations and additive intensity changes. 

\noindent
\textbf{Reformulation.} In the deep learning framework, we use the map-based image registration and disentangle the velocity fields into two parts, i.e., $v_t^{sd}$ that drives the spatial deformation $\phi$ and $v_t^{iv}$ that drives the intensity variation $q$. Hence, we replace the transport equation in Eq.~\eqref{eq:metamorphosis} with a reformulation of two separate dynamics, using the following two ordinary differential equations (ODEs):
\begin{equation}
    \frac{d \phi}{dt} = v_t^{sd} \circ \phi(t), \phi(0) = id \quad and \quad \frac{d q}{dt} = v_t^{iv} \circ q(t), q(0) = 0, 
\label{eq:deepMeta}
\end{equation}
where $id$ is the identity map, the spatial deformation $\phi$ lies in a vector space that has the same size as the velocity field $v$, and the intensity variation $q$ is scalar and has the same size as the image $I$. At $t=1$, the transported image $I(1)$ is the combination of the deformed source image and the total intensity changes. To restrict intensity changes within the pathological regions, we use its mask $\mathcal{M}$ and obtain
    $I(1) = \phi(1) \circ I_0 + q(1) \otimes \mathcal{M}$.
This formulation models the appearance of a tumor from a clean source image to a pathological target image.


\section{Deep Metamorphic Image Registration}
Based on the above reformulation, we propose a metamorphic image registration network (MetaRegNet) to jointly model spatial deformations and intensity changes. Overall, MetaRegNet adopts a UNet network to estimate two initial velocity fields $v_0^{sd}$ and $v_0^{iv}$, which drive the spatial deformation $\phi$ and the intensity variations $q$ over time. The combination of the deformed source image and estimated intensity changes matches the target image. The proposed architecture is presented in Fig.~\ref{fig:metaregnet}, with each component described below.   



\noindent
\textbf{Estimation of Initial Flows.} To obtain spatial deformations and intensity variations, we first estimate their initial driven flows. We adopt a UNet~\cite{ronneberger2015u} to estimate the initial values $v_0^{sd}$ and $v_0^{iv}$ from an image pair. As shown in Fig.~\ref{fig:metaregnet}, the network includes a non-probabilistic U-Net architecture, which directly outputs the flow estimation, without sampling from the mean and variance as in~\cite{dalca2018unsupervised}. This non-probabilistic approach is simpler and works better in our experiments.



\noindent
\textbf{Integration of Spatial Deformations and Intensity Variations.} To solve Eq.~\eqref{eq:deepMeta} with the estimated initial flows, we utilize Lipschitz Continuous ResNet Blocks (LC-RB) proposed in~\cite{joshi2021diffeomorphic}. These LC-RB blocks are used as numerical integration schemes for solving ODEs. We use the version without sharing weights among seven blocks, which models time-varying velocity fields and produces diffeomorphic deformations. 
That is, we obtain the diffeomorphic deformation $\phi(1)$, which captures the spatial transformations between the source and target images and deforms the healthy source image to generate parts of the final image. 

Another branch with the same number of LC-RB blocks produces incremental residual mappings of additive intensity changes between the input pair, starting from $q_{0}$. The intensity variations produced at the end, $q(1)$, are added onto the deformed source to approximate the target image $I_1$. Naively, if $q(1)$ was a simple pixel-/voxel-wise subtraction of the input pair, it would perfectly reconstruct the target image; however, in such a case, the anatomical matching constraint of image registration will not be met. 
To avoid this issue, we add a regularization similar to~\cite{maillard2022deep}, which restricts the learned intensity variations within the pathological region of the target image. That is, missing correspondences only happen within the pathological regions. We assume the availability of the binary mask $\mathcal{M}$ of the pathological region, like a tumor, and use it to mask out intensity variations in non-pathological regions, i.e., $q(1) \otimes \mathcal{M}$.

\noindent
\textbf{Interpolation and Output.} In this step, we generate the final output to approximate the target image $I_1$. Firstly, with $\phi(1)$ we generate the deformed source image using a differentiable interpolation layer. For each voxel $p$ in the target image, this layer computes its location given at $\phi(p)$ in the source image and obtains its intensity value using linear interpolation. Upon this, we produce our final metamorphic output by adding the generated intensity variations, which performs a pixel-wise addition of $q(1) \otimes \mathcal{M}$ to the deformed image $\phi(1)\circ I_0$.

\noindent
\textbf{Loss Function.} Similar to metamorphic image registration formulated in Eq.~\eqref{eq:metamorphosis}, we have the image matching between our metamorphic output and the target image, and the regularization on the spatial deformation and the intensity variation. The overall loss function is formulated as
\begin{align*}
   \mathcal{L} = & \lambda_{1}\mathcal{L}_{sim}(\frac{1}{|\Omega|} {\|I_{1} - \phi \circ I_{0} \|}_2^2) +  \lambda_{2}\mathcal{L}_{reg}(\nabla (q(1)\otimes \mathcal{M}) \\ 
     + & \lambda_{3}\mathcal{L}_{Jdet}(0.5(|\mathbb{J}(\phi(1))| - \mathbb{J}(\phi(1)))),
\end{align*}
where $\Omega$ is the image spatial domain and $|\Omega|$ indicates the number of pixels or voxels in an image, and $\lambda_{1}$, $\lambda_{2}$ and $\lambda_{3}$ are the balancing weights, which are set to $[1.0, 1.0, 0.001]$, respectively, in our experiments.
We use mean squared error (MSE) as the image similarity metric $\mathcal{L}_{sim}$ to measure the goodness of image matching. We also discourage dramatic intensity changes within the learned intensity values by using a diffusion regularizer $\mathcal{L}_{reg}$ on the estimated intensity changes, where $\nabla$ is the spatial gradient operator. Besides, to restrict the learned deformations to be diffeomorphic we use $\mathcal{L}_{Jdet}$, where we penalize the total number of locations where the Jacobian determinants $|\mathbb{J}(\phi(1))|$ are negative.


\section{Experiments}

\noindent
\textbf{Datasets.} 
{\it (1) BraTS 2021~\cite{baid2021rsna,menze2009menze,kuzilek2017open}.} This dataset includes image scans collected from 1251 subjects. Each scan is pre-processed by skull-strpping, co-registering to a common anatomical template, and being interpolated to the same resolution of $1 \times 1 \times 1 mm^{3}$, which is followed by an intensity normalization between $0$ to $1$. We select 120 slices with no tumor as our healthy image set and 120 slices with tumors as our pathological set, by checking their corresponding tumor masks. We keep aside 20 images from each set for testing and 5 for validation. As a result, we have 240 random image pairs for training, 5 pairs for validation, and 20 pairs for testing. 
{\it (2) 3D-IRCADb-01~\cite{3dircadb}.} This database is composed of 3D CT scans of 20 different patients with hepatic tumors. Each image has 74$\sim$260 slices with size of $512 \times 512$, which is resampled to a pixel spacing of 1mm. Since the pixel values are in Hounsfield and in the range of $[-1000,4000]$, we perform a color depth conversion using the mapping as in~\cite{almotairi2020liver}: $g = \frac{h - m_{1}}{m_{2} - m_{1}} \times 255$. Here, $g$ is the converted gray level value, $h$ is the Hounsfield value in the raw image, and $m_{1}$ and $m_{2}$ are the minimum and maximum of the Hounsfield range, respectively. Then, we crop the liver region using the provided liver mask and normalize the image intensity to [0, 1].
We collect 20 slices that contain healthy liver regions and 6 slices that have a pathological regions in the liver. We take 12 slices from the healthy set and 3 slices from the unhealthy set to make our training set, resulting in 36 image pairs, and take 3 healthy slices and 1 unhealthy slice to make 3 pairs for validation, and 5 healthy slices and 2 unhealthy slices to make 10 pairs for testing. 
Due to the limited training samples, we extend the training set to 144 pairs, via rotating each image by 90, 180 and 270 degrees. For both datasets, we have {\it subject-wise} splitting for training, validation, and test sets.   


\noindent
\textbf{Baseline Methods.} For comparison, we choose the diffeomorphic version of VoxelMorph~\cite{dalca2018unsupervised}, a deep-learning-based image registration model, as our baseline. Since VoxelMorph cannot handle the metamorphic image registration problem, we modify it and adopt the cost-function-masking (CFM) strategy~\cite{brett2001spatial} to exclude the similarity measure of the tumor regions using their masks during training. This modified VoxelMorph is denoted as VM-CFM.


\noindent
\textbf{Implementation and Settings.} We implement our MetaRegNet using Keras and TensorFlow and train it in an end-to-end fashion, with the Adam optimizer and a fixed learning rate of $1e^{-4}$. Both our architecture and the baseline model VM-CFM have been deployed on the same machine with an Nvidia TITAN X GPU. We build our method on top of the R2Net implementation with default parameters reported in~\cite{joshi2021diffeomorphic}. All models are trained from scratch. 



\noindent
\textbf{Evaluation Metrics.} We measure the average Sum of Squared Distance (SSD) between the deformed source and target images, including the whole image (SSD-total) and the healthy region only (SSD-healthy). 
To measure the number of foldings in the estimated spatial deformations, we report the number of voxels with negative Jacobian determinants. Also, we measure the segmentation Dice score by using estimated deformations and the inference time as well.


\begin{figure}[t]
\centering
\setlength{\tabcolsep}{0.5pt}
    \begin{tabular}{lllllll}
         \rotatebox[origin=l]{90}{\scriptsize Image Pair} &
         \includegraphics[width=0.14\textwidth]{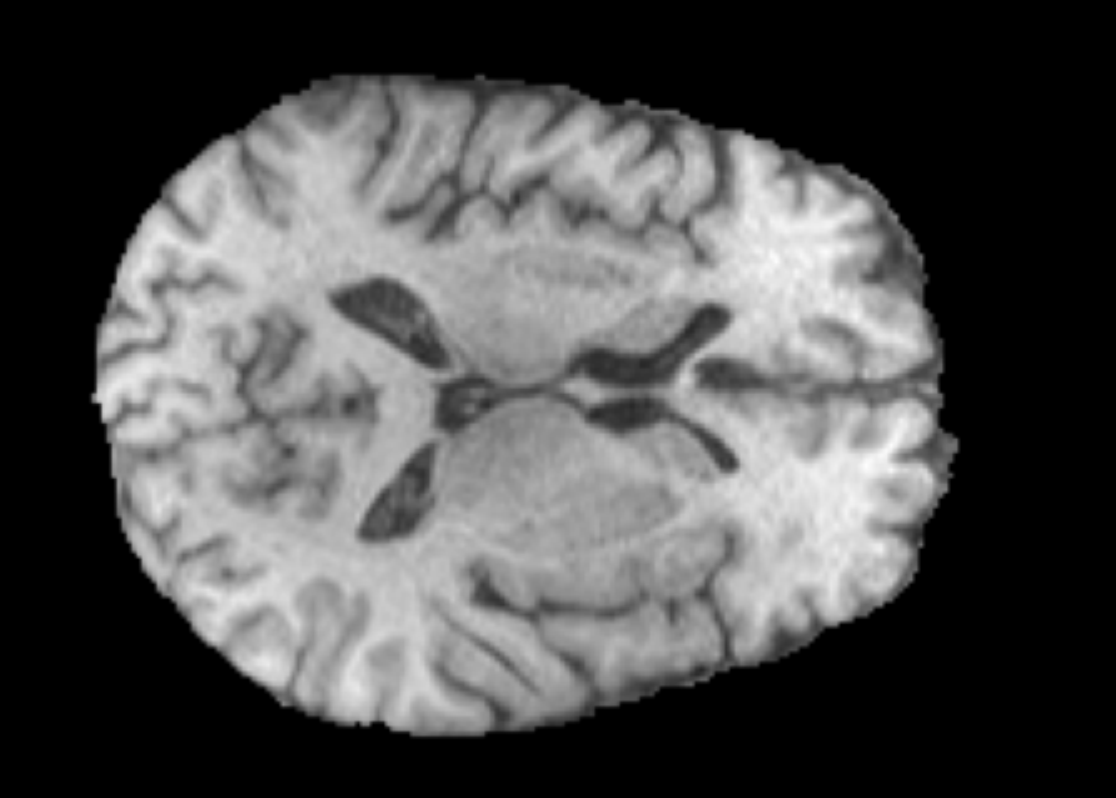}  &
         \includegraphics[width=0.14\textwidth]{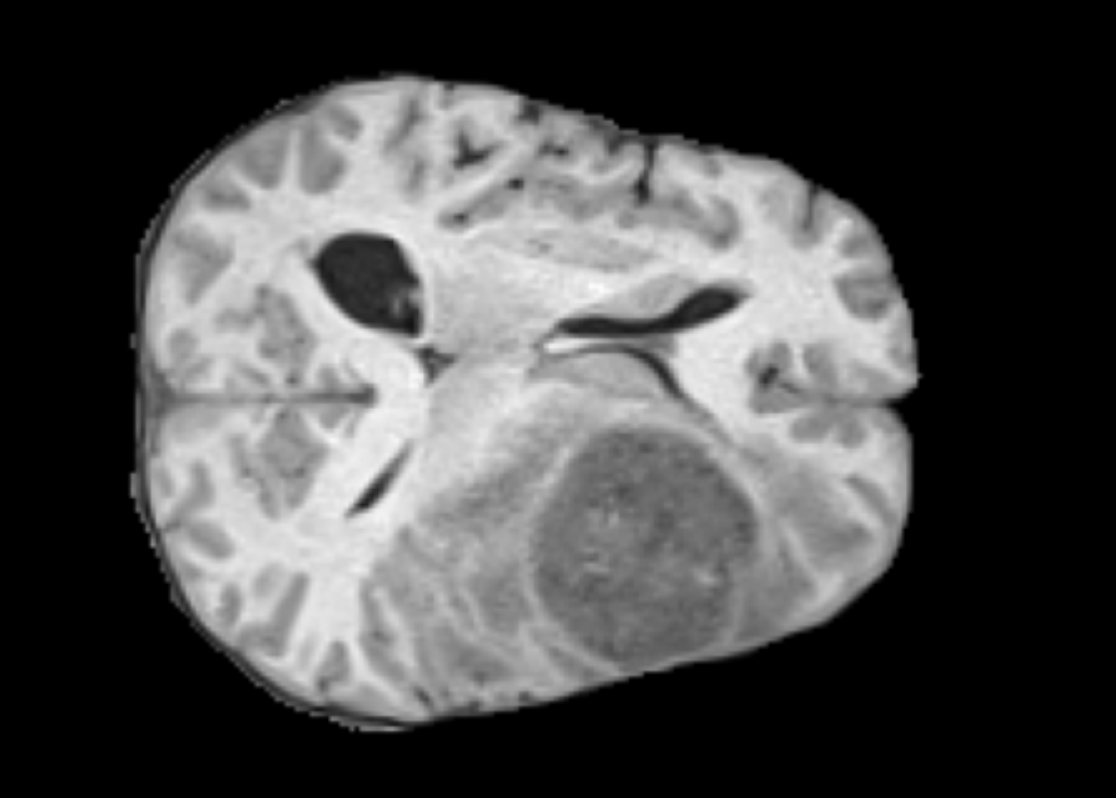} &
         &
         \;
         \includegraphics[width=0.14\textwidth]{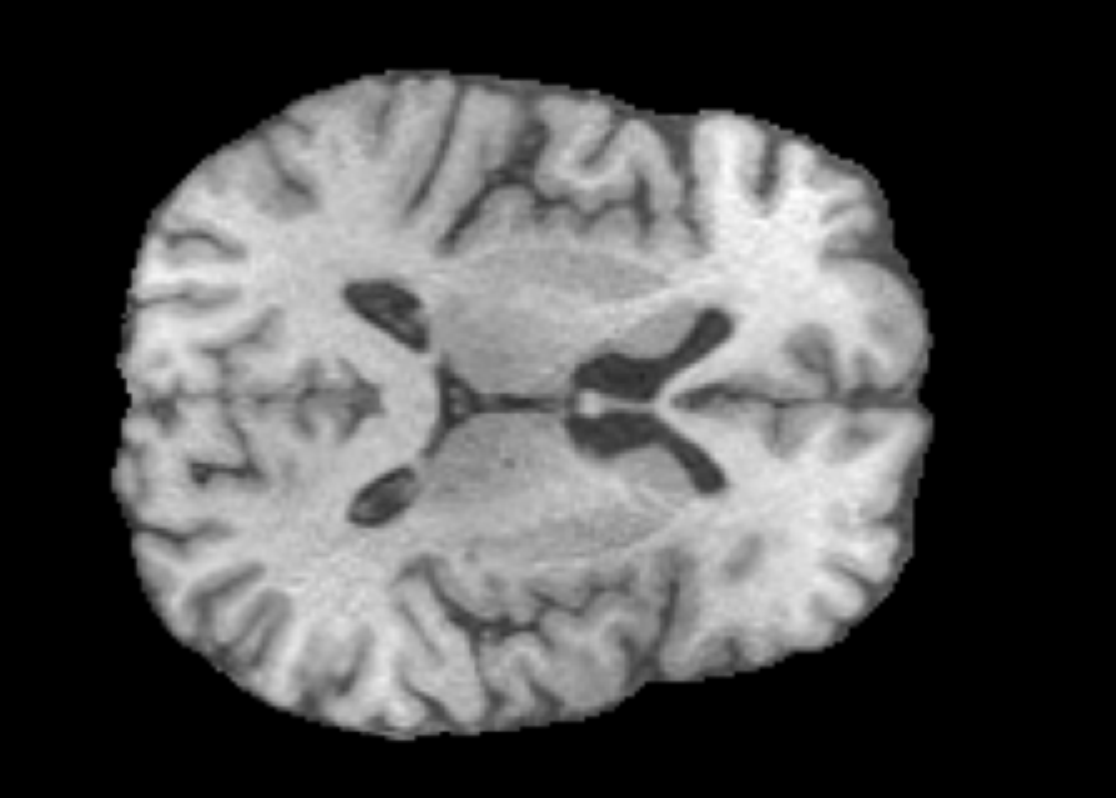}  & \includegraphics[width=0.14\textwidth]{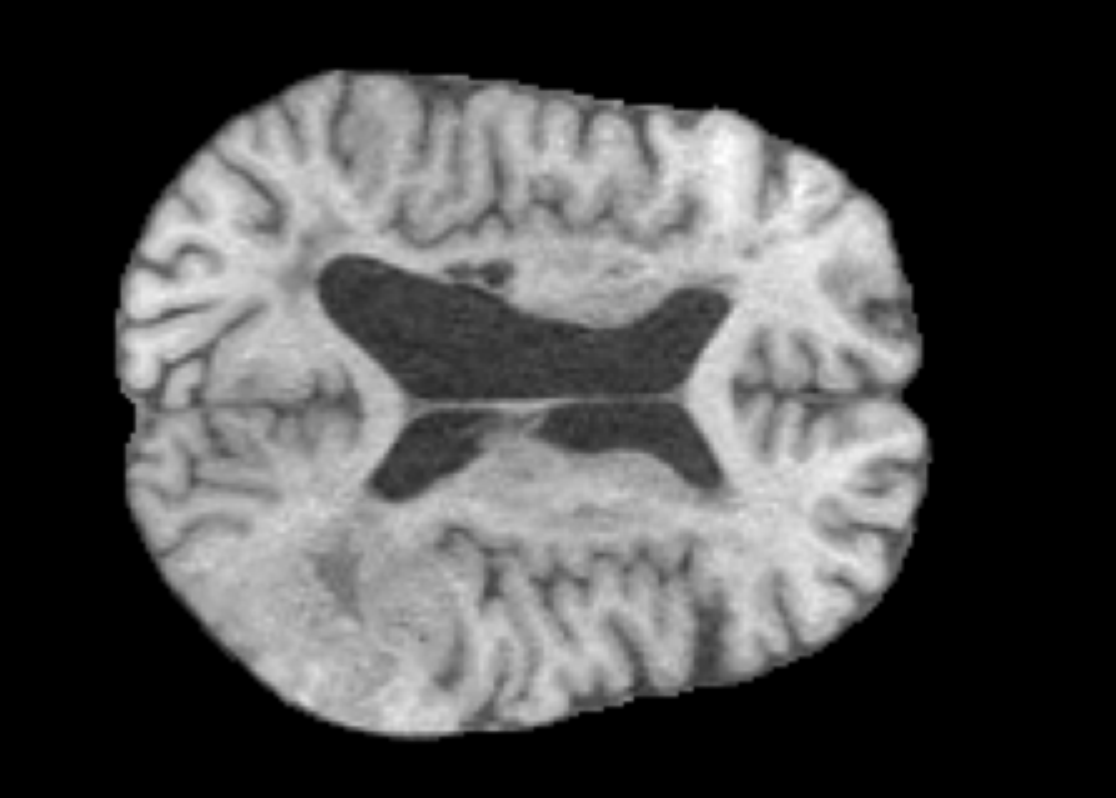}  & \\
         \rotatebox[origin=l]{90}{\scriptsize VM-CFM} &
         \includegraphics[width=0.14\textwidth]{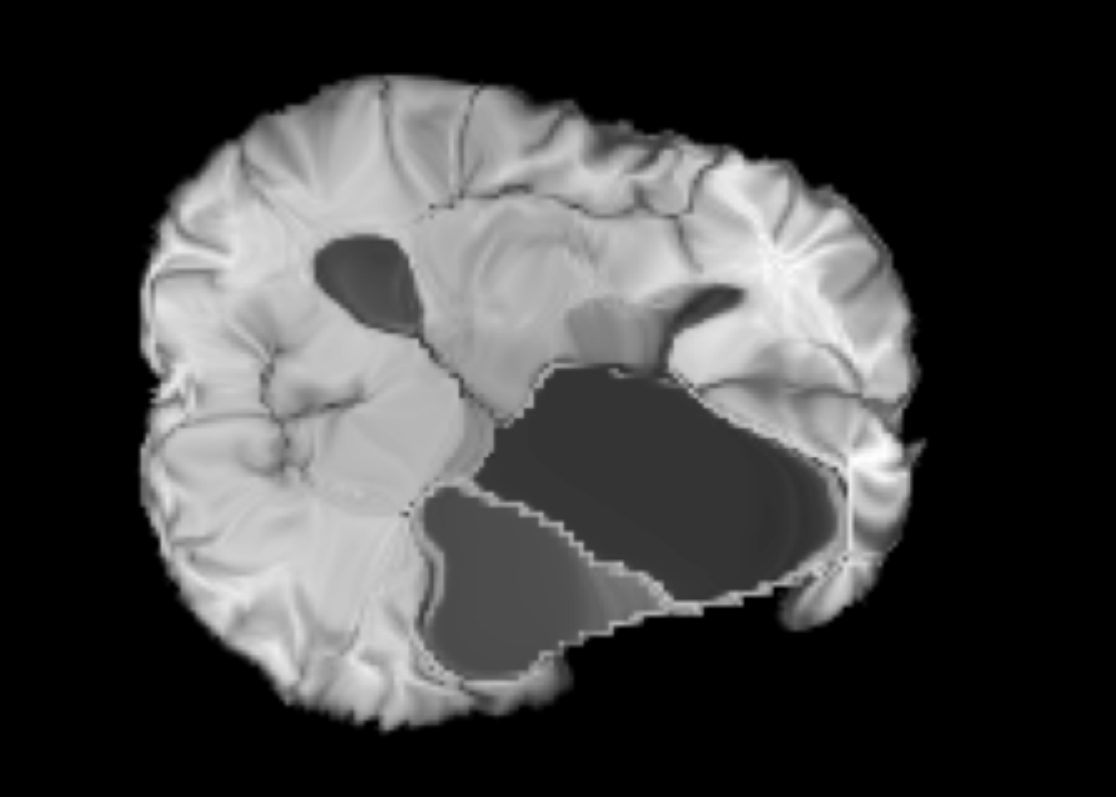} 
         &
        \includegraphics[width=0.14\textwidth]{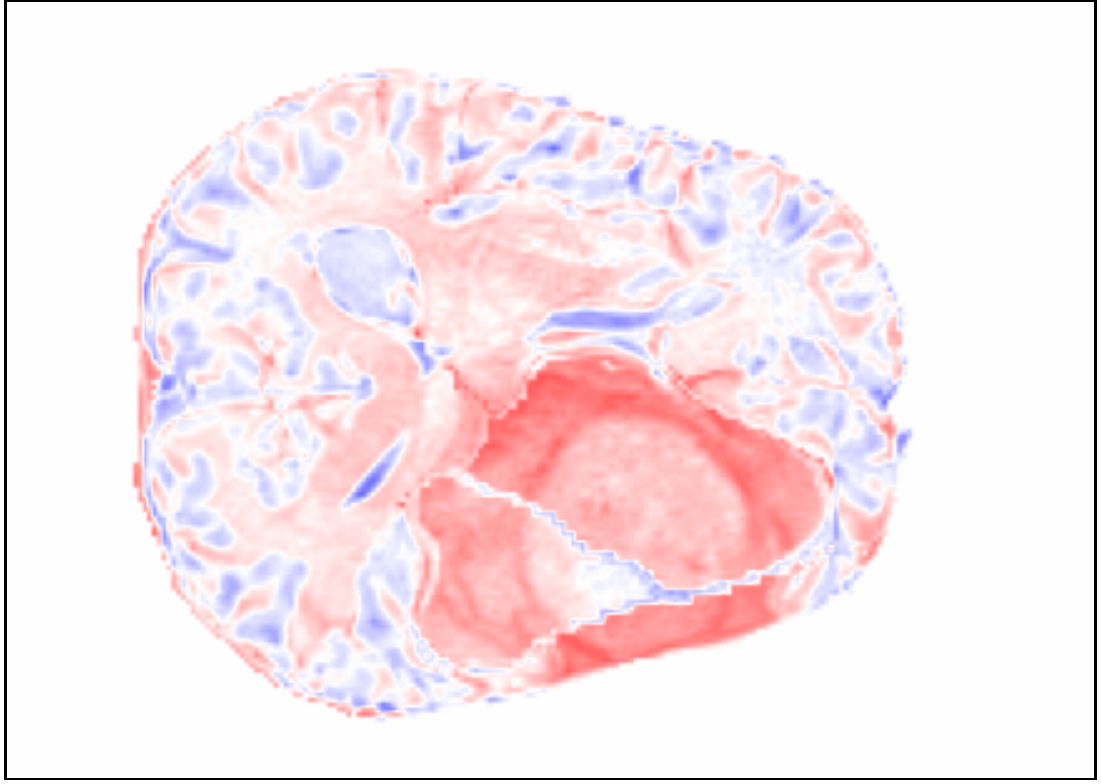} &
        \includegraphics[width=0.14\textwidth]{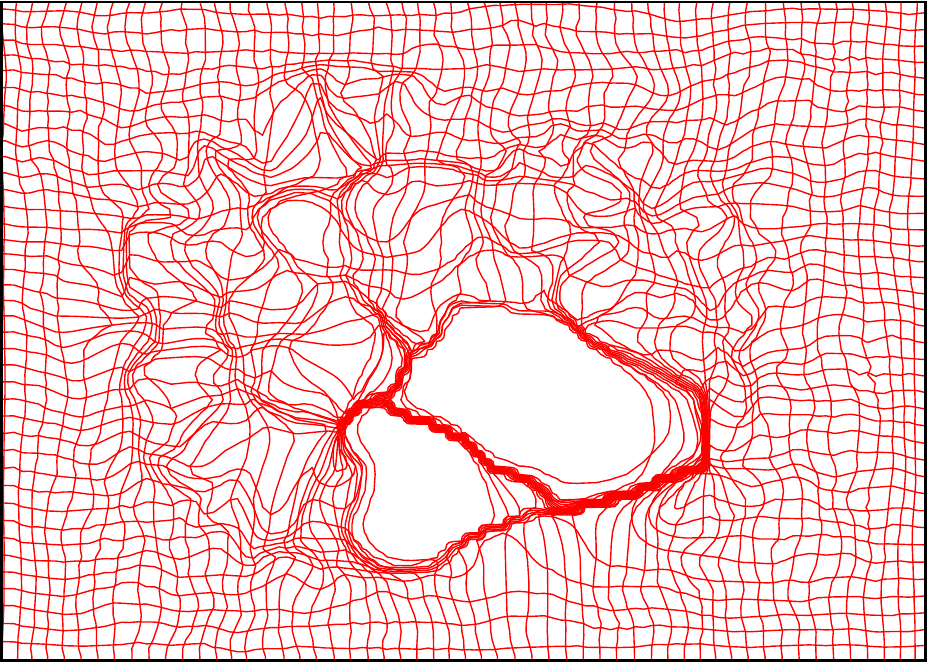} & \;
        \includegraphics[width=0.14\textwidth]{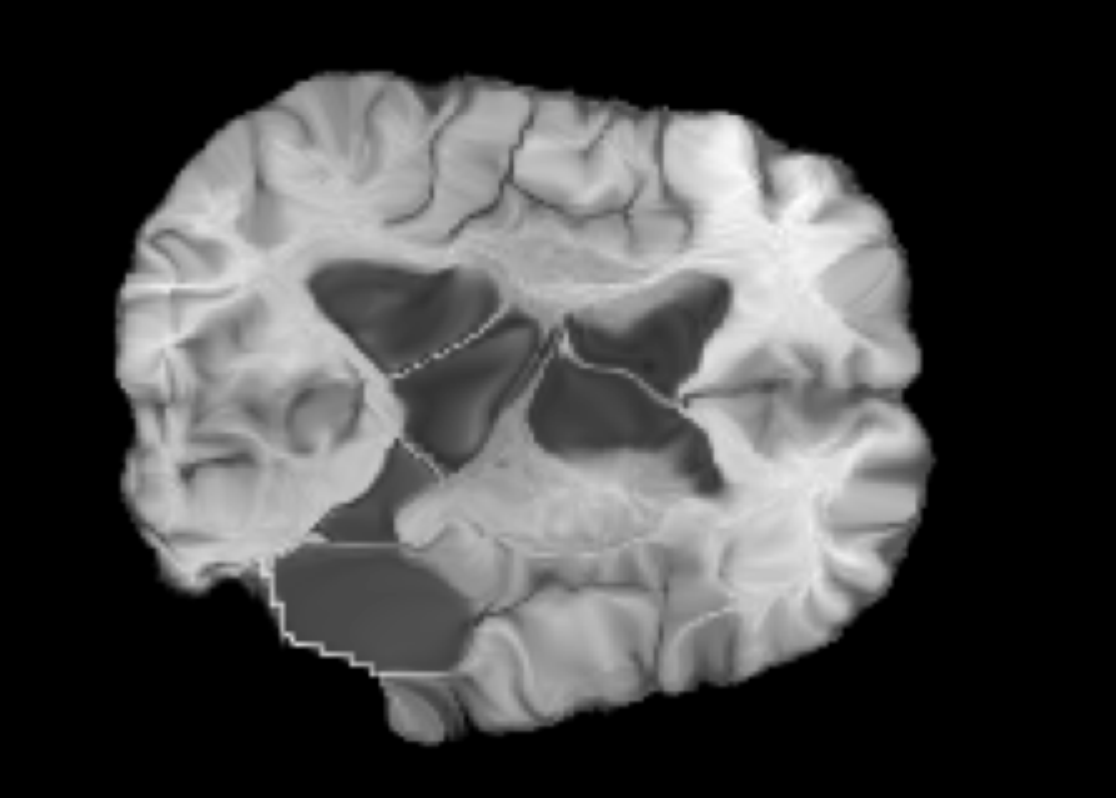}     &
        \includegraphics[width=0.14\textwidth]{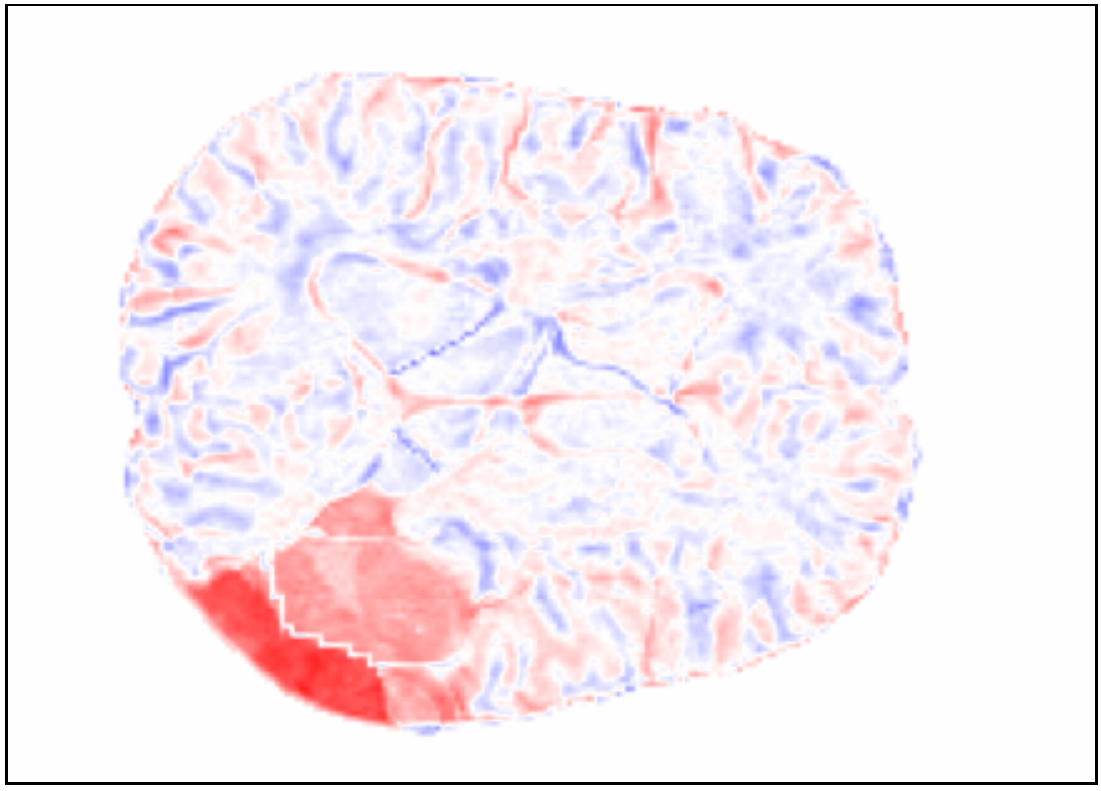} &
        \includegraphics[width=0.14\textwidth]{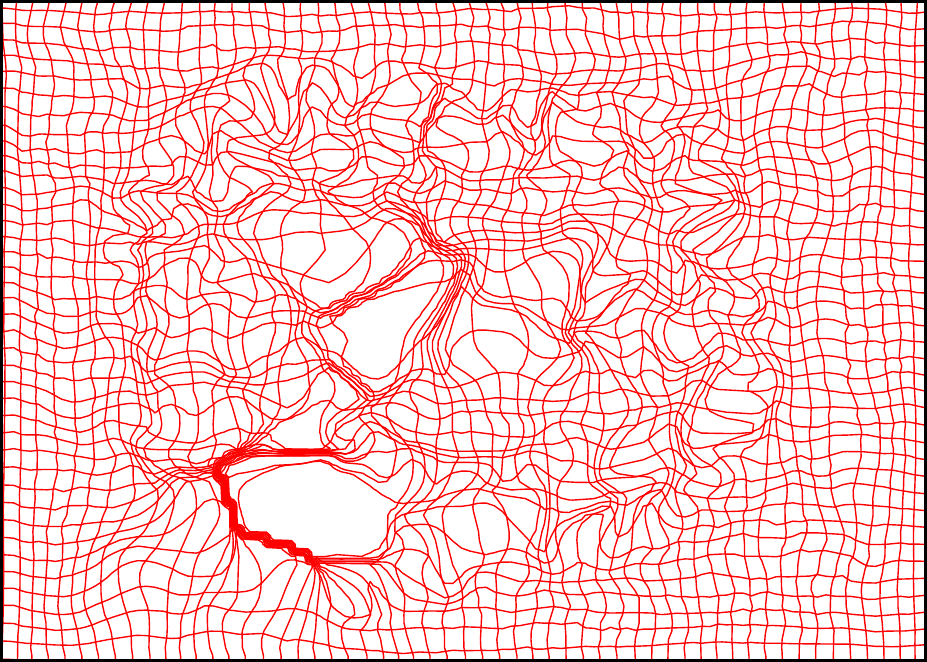}
        \\
        \rotatebox[origin=l]{90}{ \scriptsize Ours} &
        \includegraphics[width=0.14\textwidth]{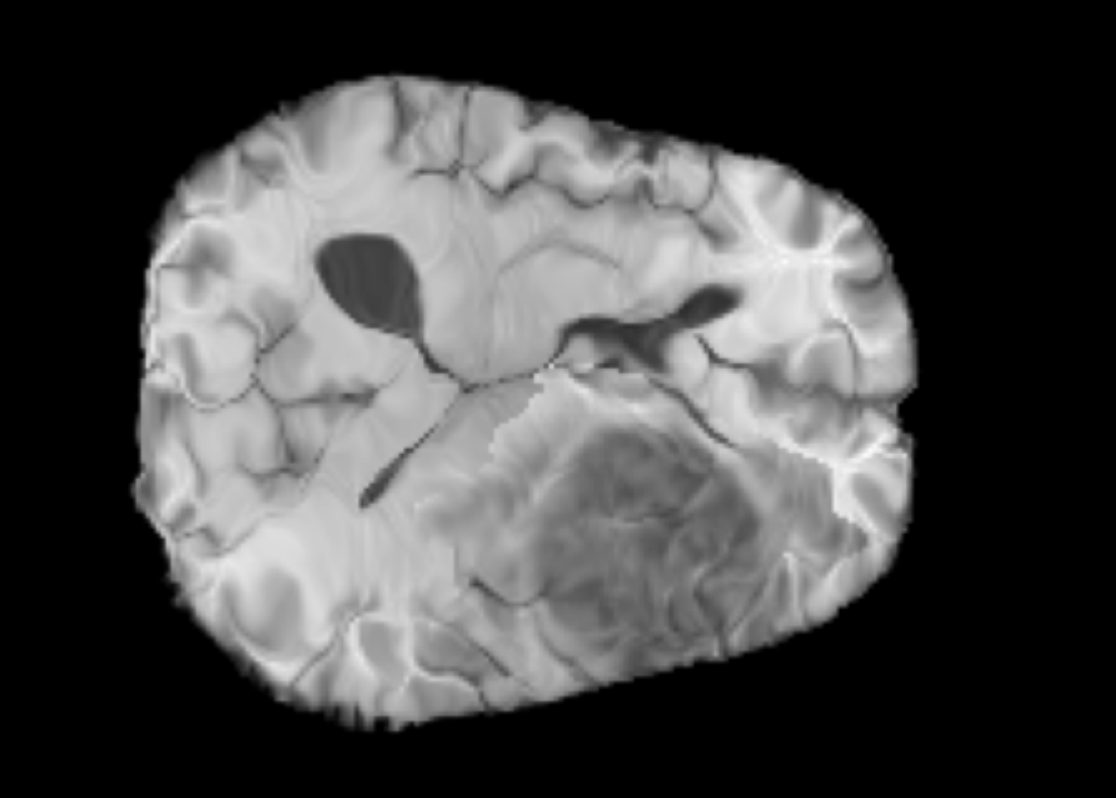}    &
        \includegraphics[width=0.14\textwidth]{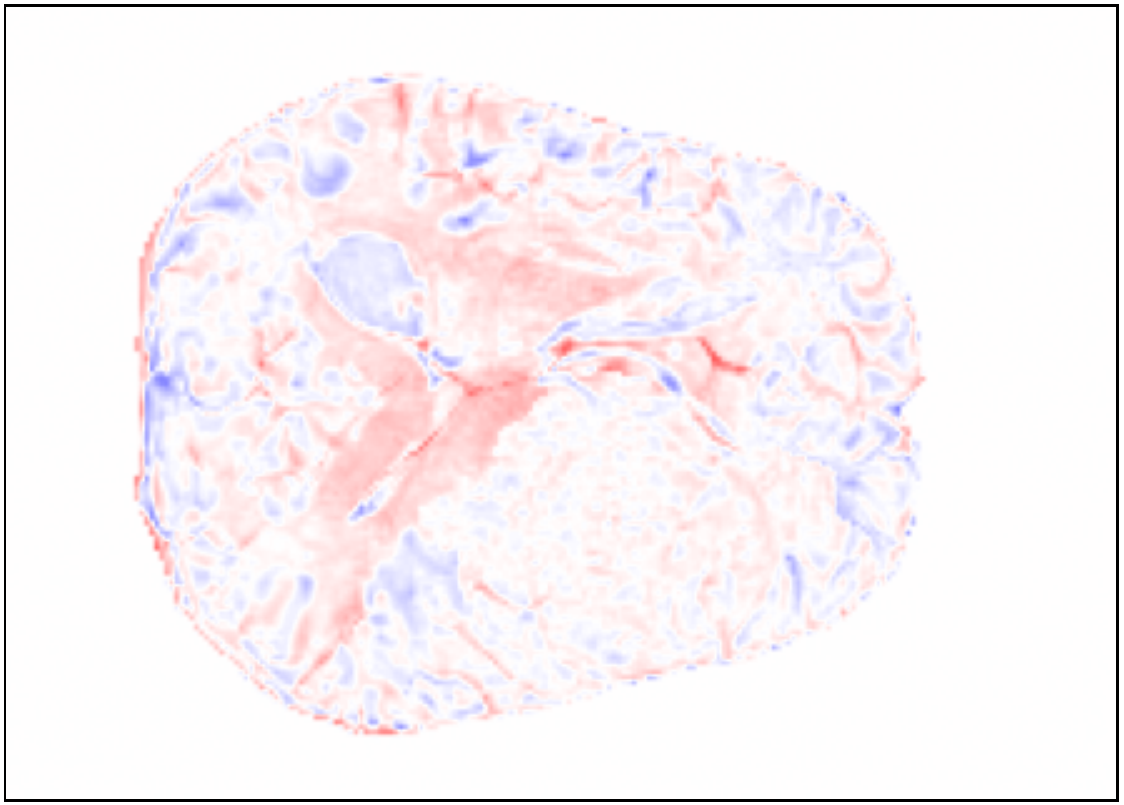} &
        \includegraphics[width=0.14\textwidth]{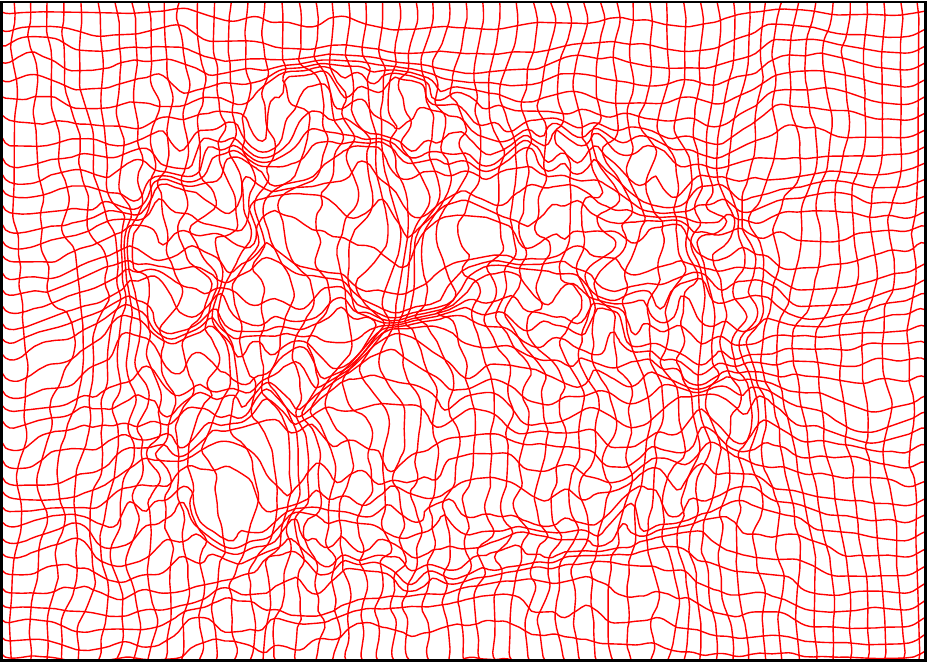} & \;
        \includegraphics[width=0.14\textwidth]{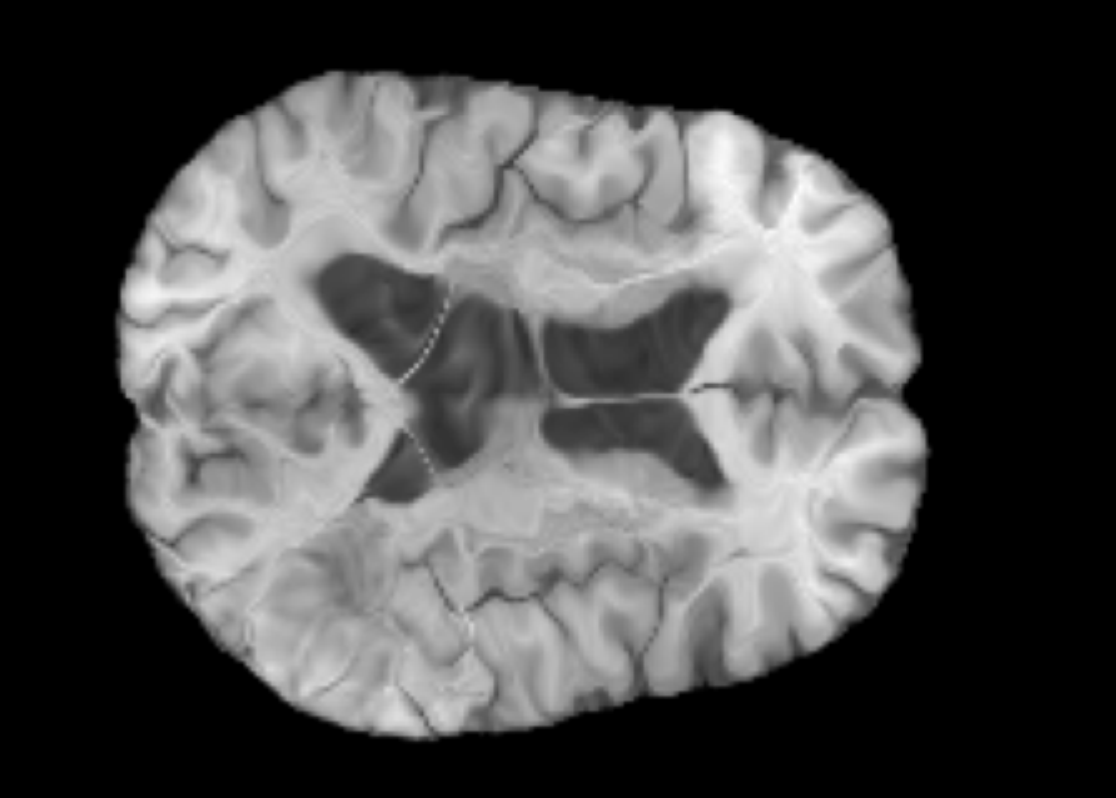}    &
        \includegraphics[width=0.14\textwidth]{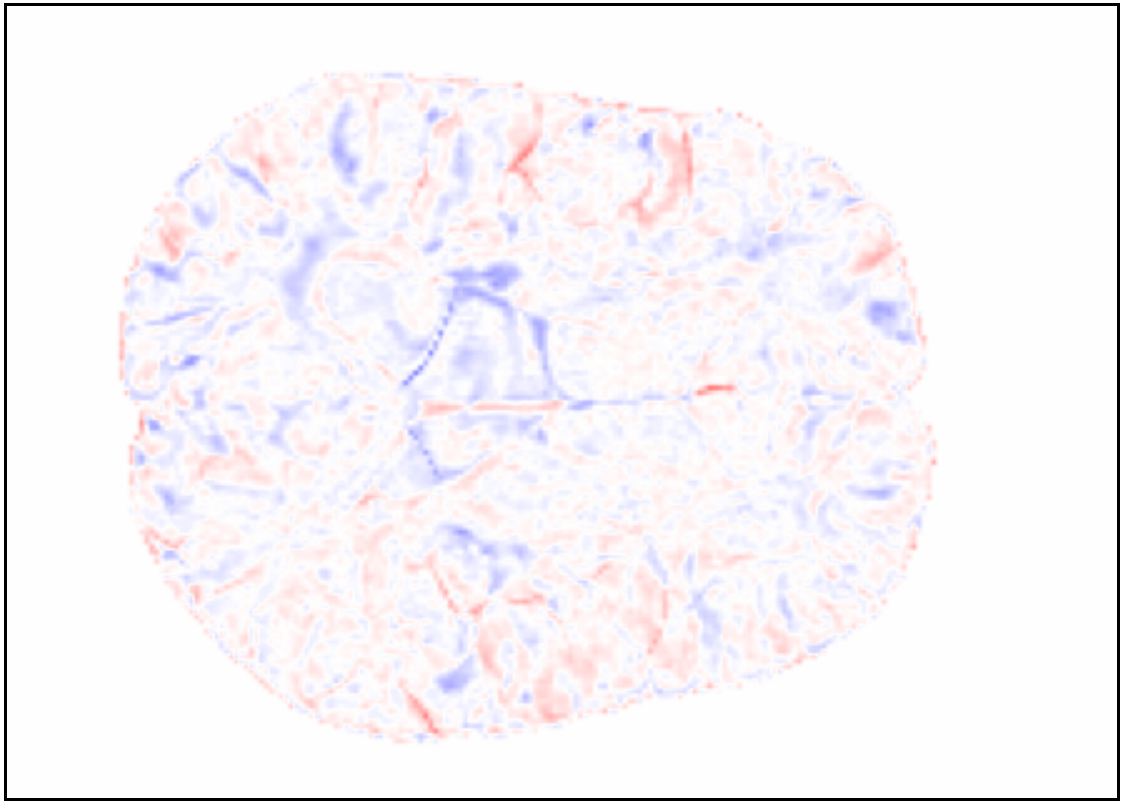} &
        \includegraphics[width=0.14\textwidth]{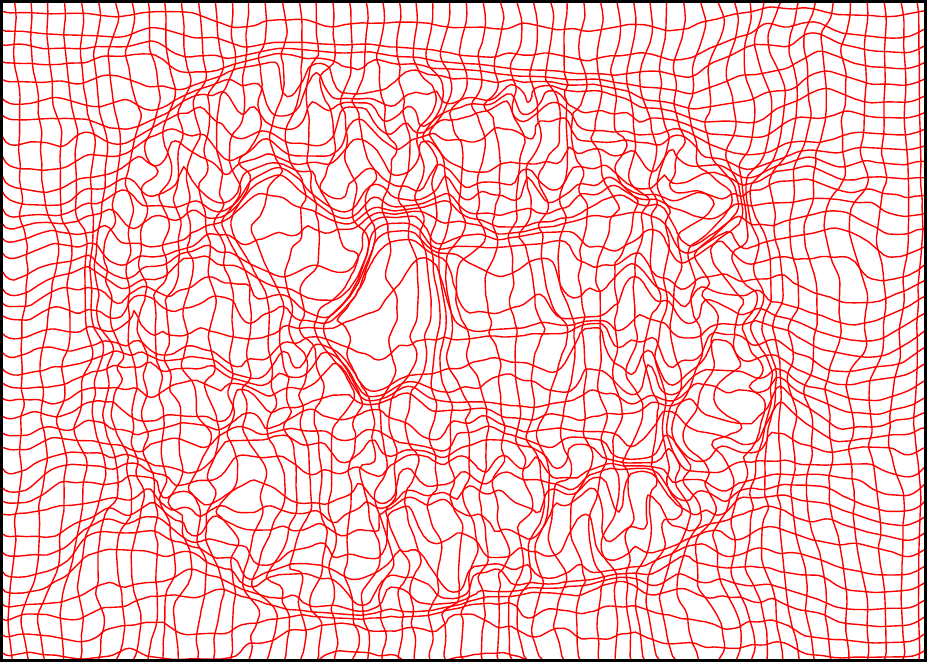} \\
         \rotatebox[origin=l]{90}{\scriptsize Image Pair} &
         \includegraphics[width=0.14\textwidth]{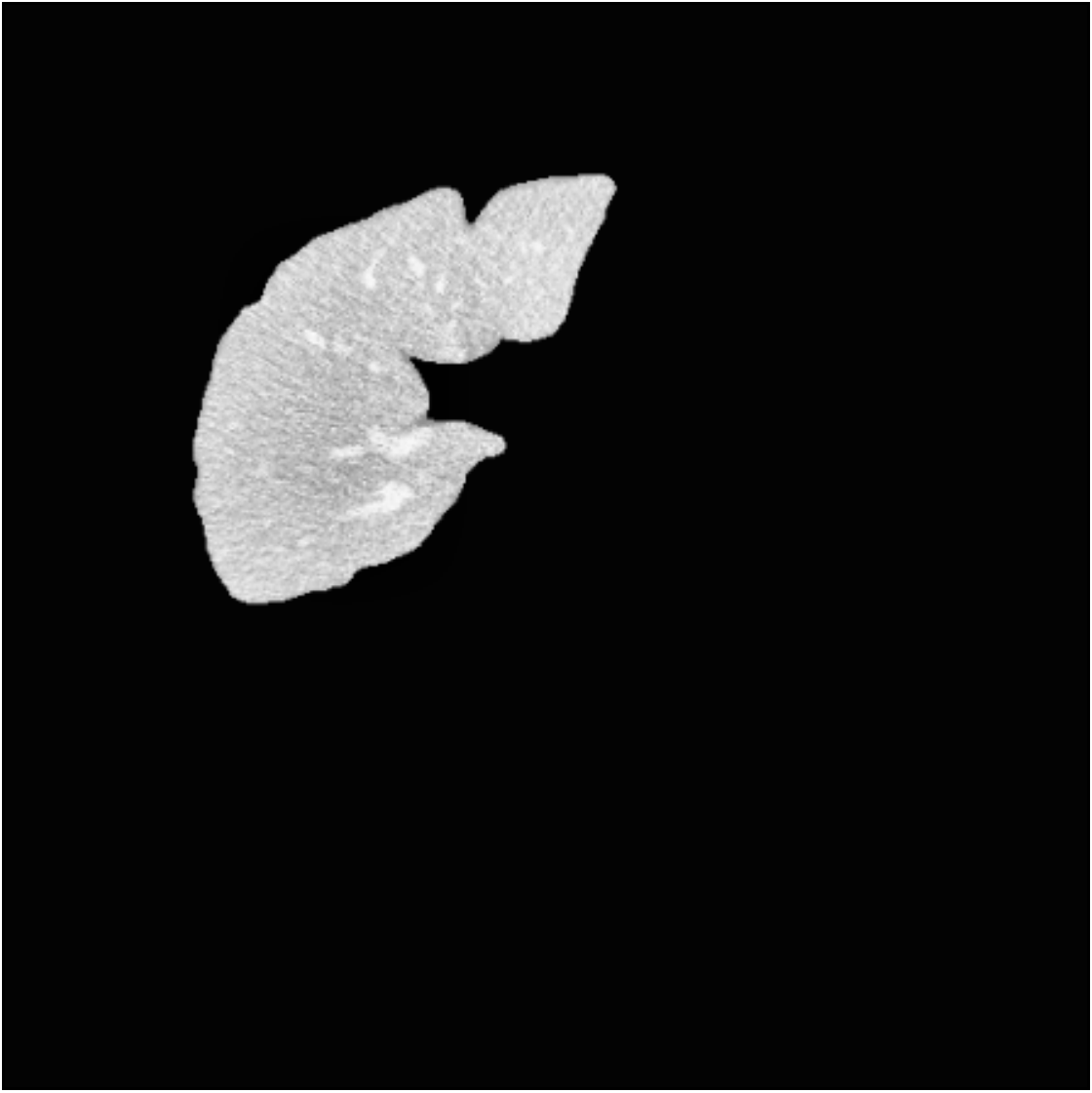}  &
         \includegraphics[width=0.14\textwidth]{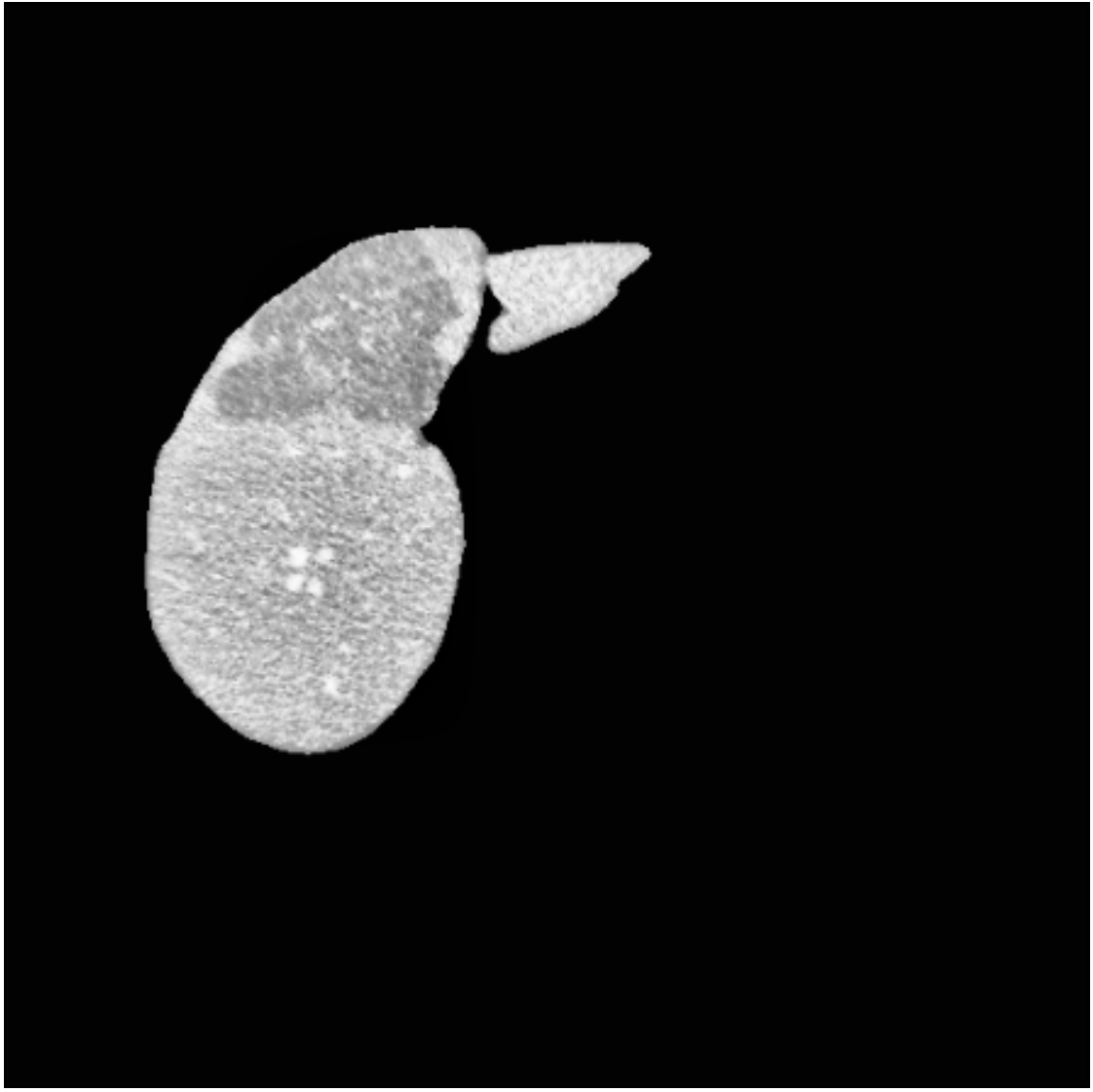} & & \;
         \includegraphics[width=0.14\textwidth]{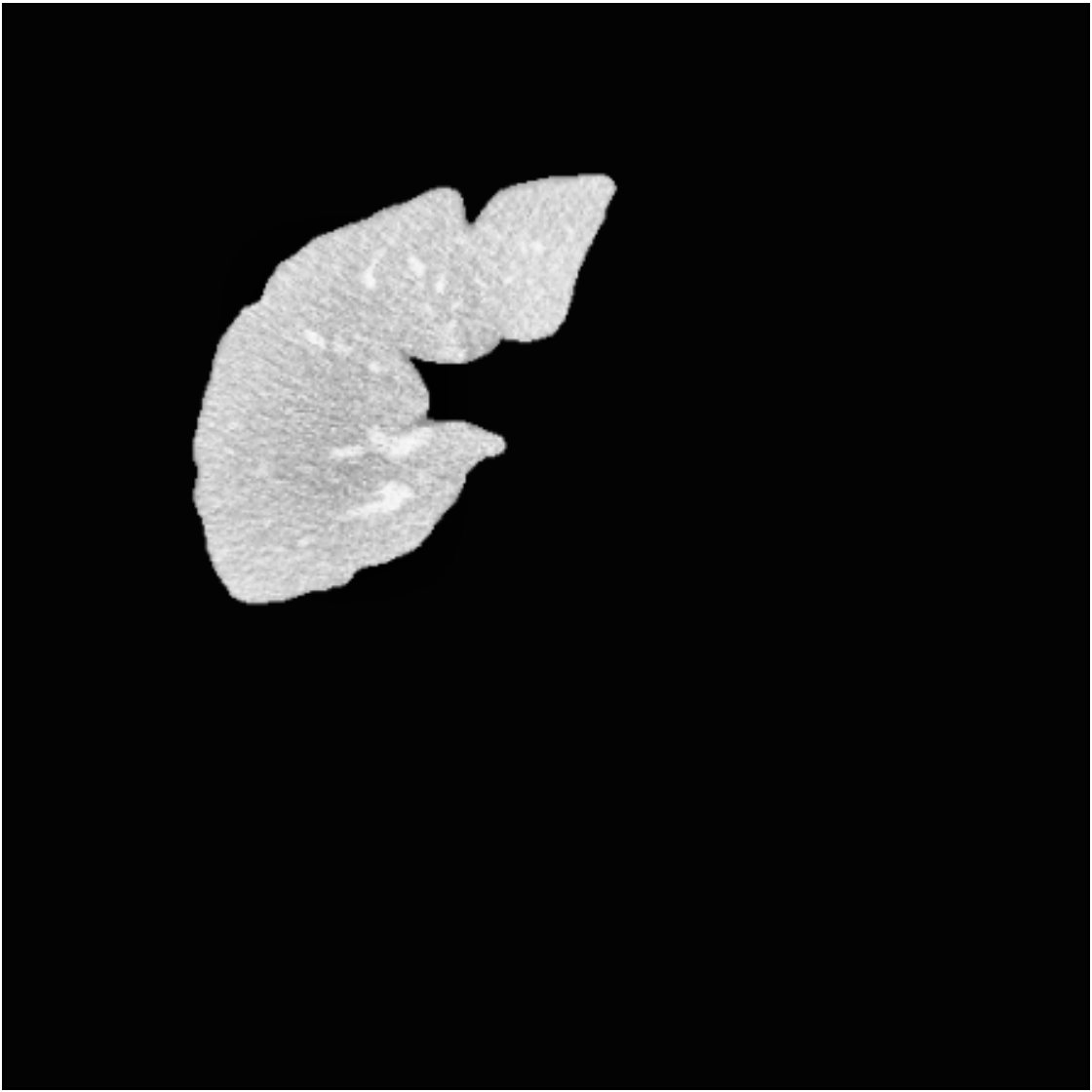}  & \includegraphics[width=0.14\textwidth]{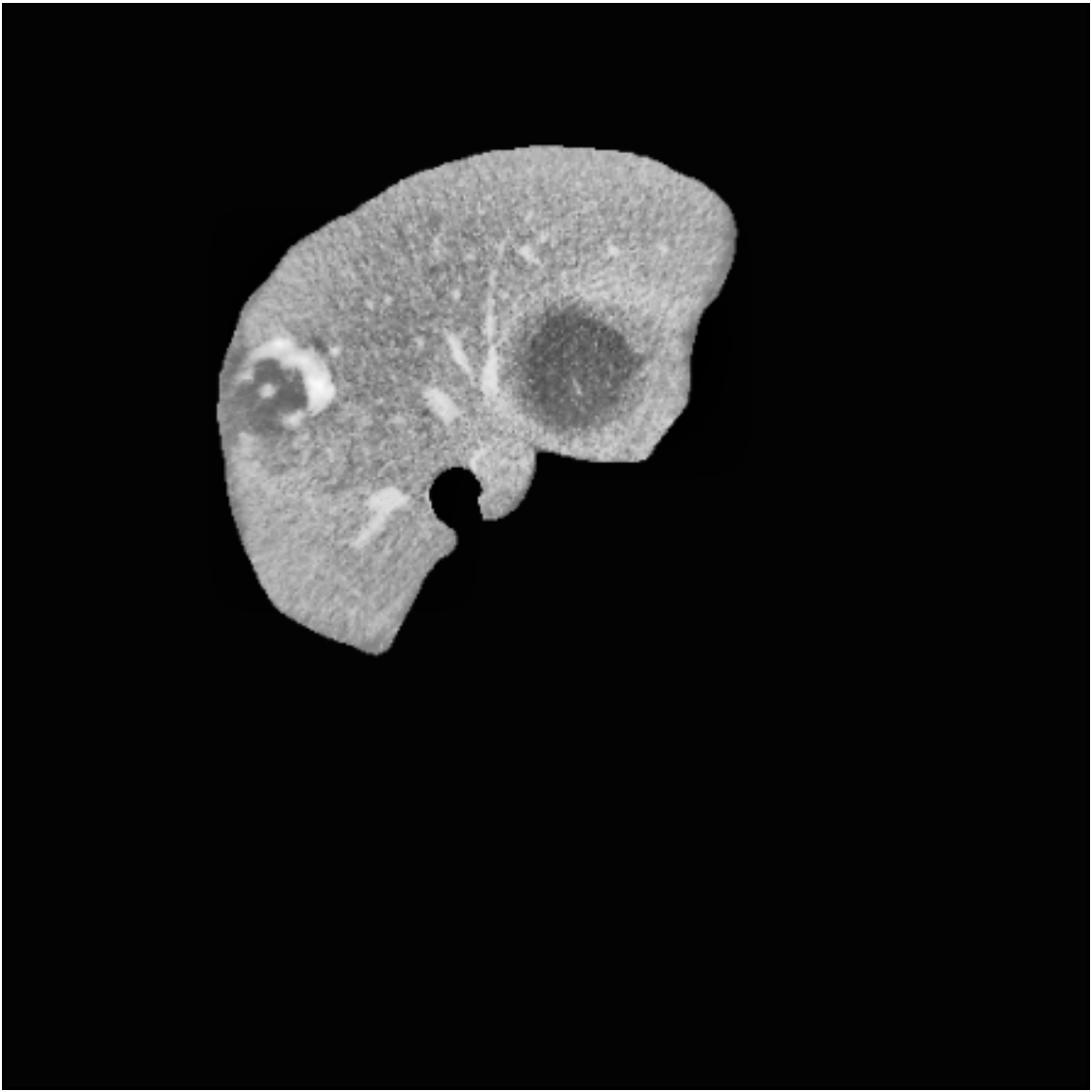}  & \\
         \rotatebox[origin=l]{90}{\scriptsize VM-CFM} &
         \includegraphics[width=0.14\textwidth]{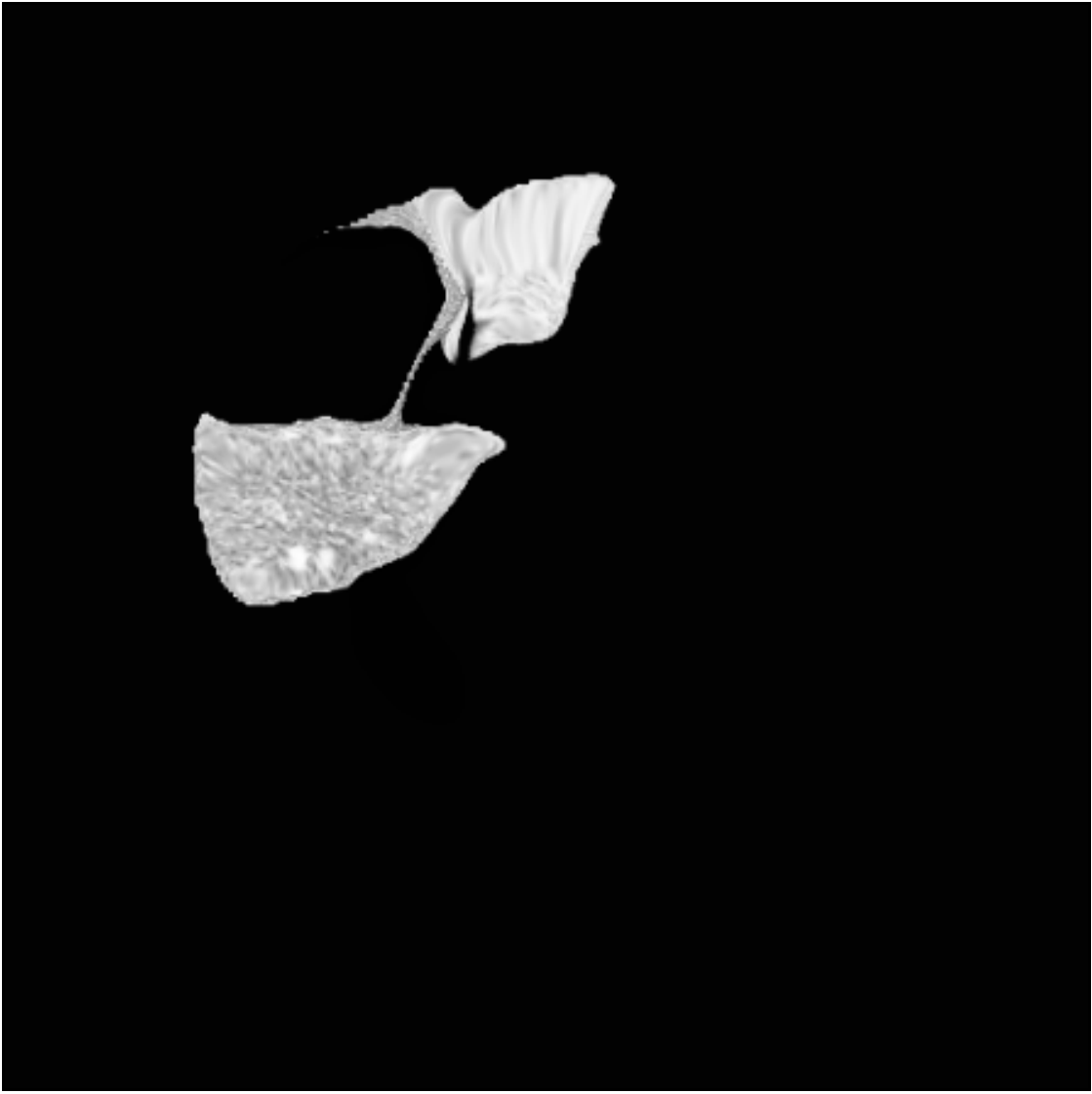} 
         &
        \includegraphics[width=0.17\textwidth]{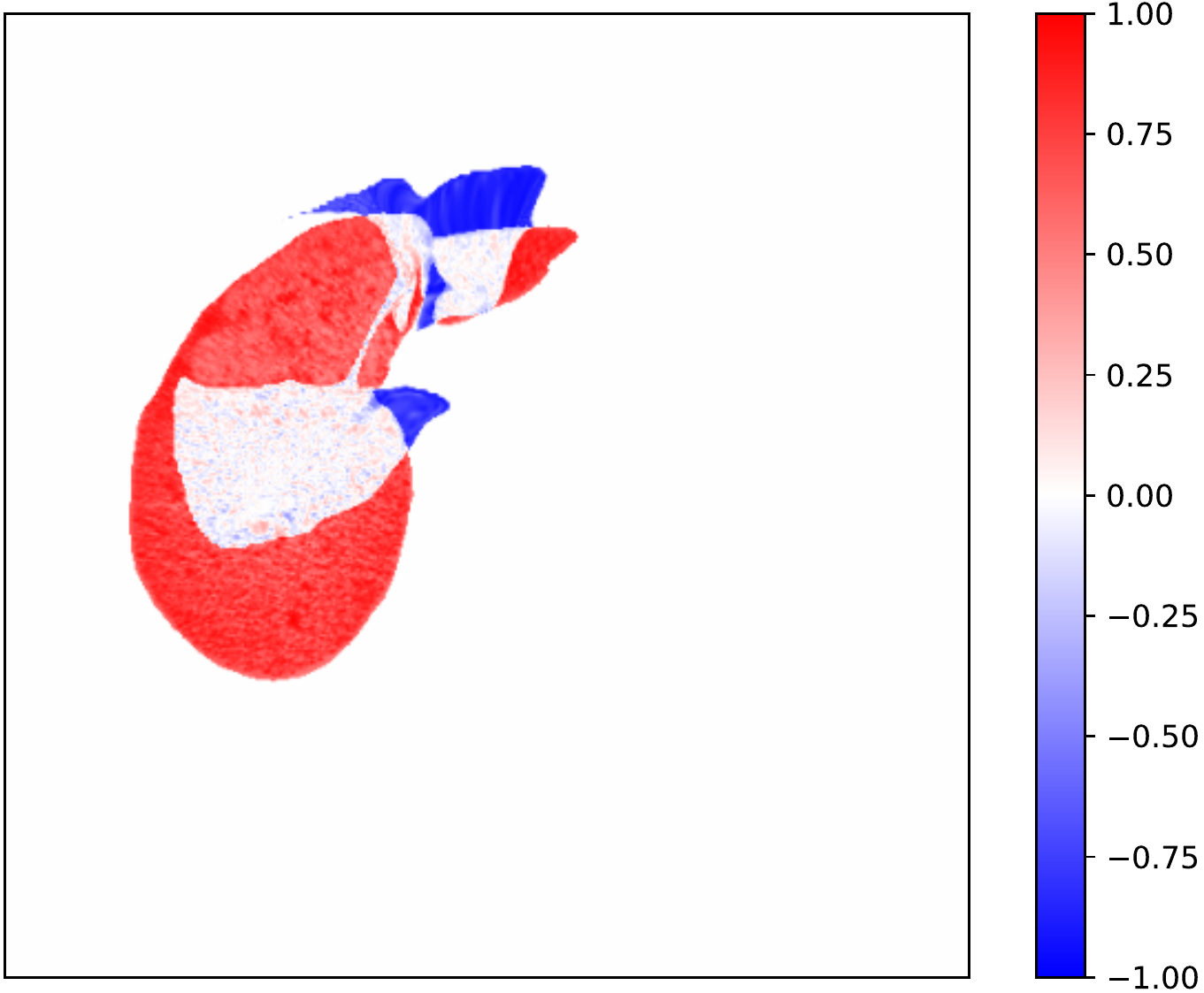} &
        \includegraphics[width=0.14\textwidth]{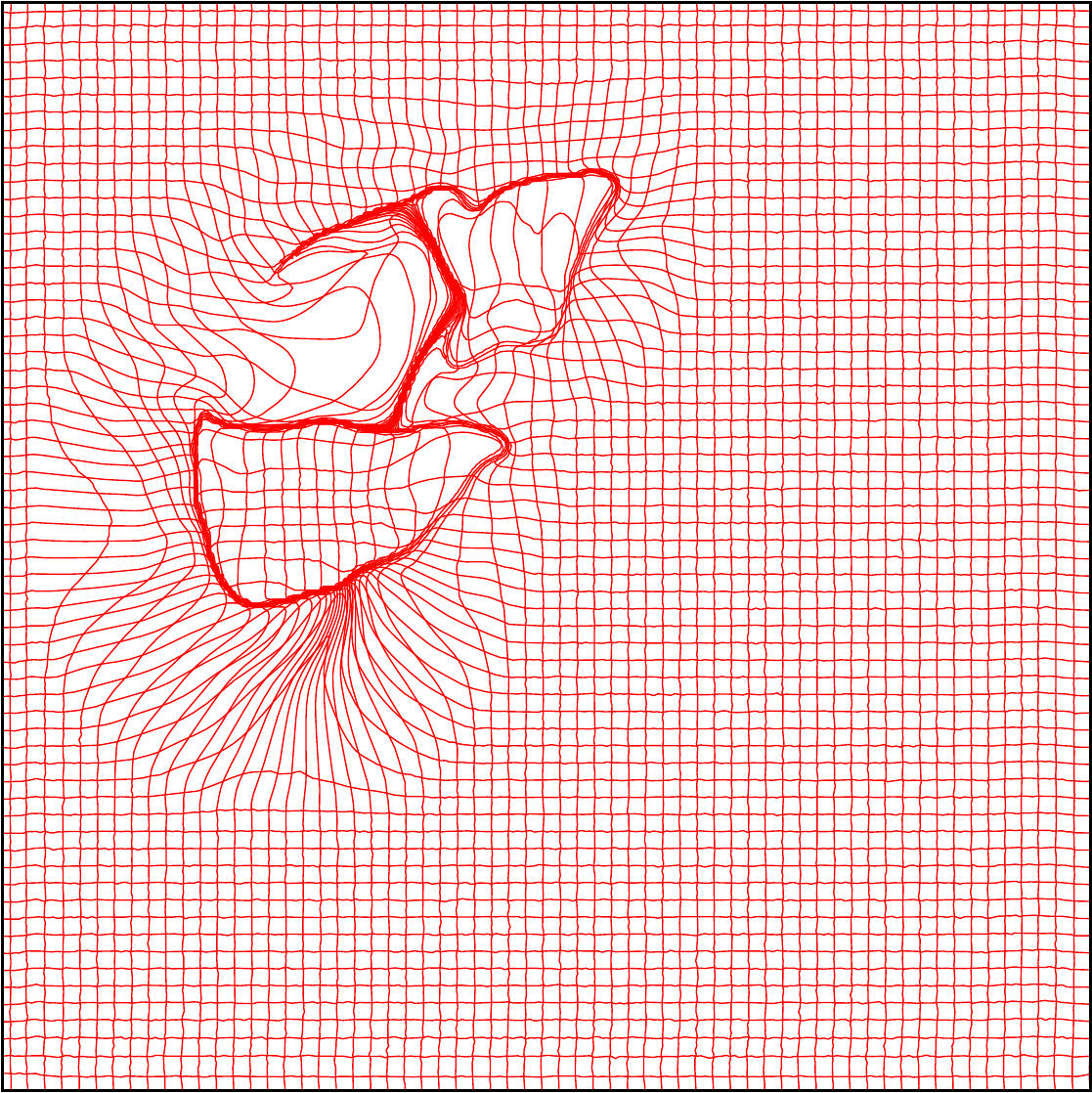} & \;
        \includegraphics[width=0.14\textwidth]{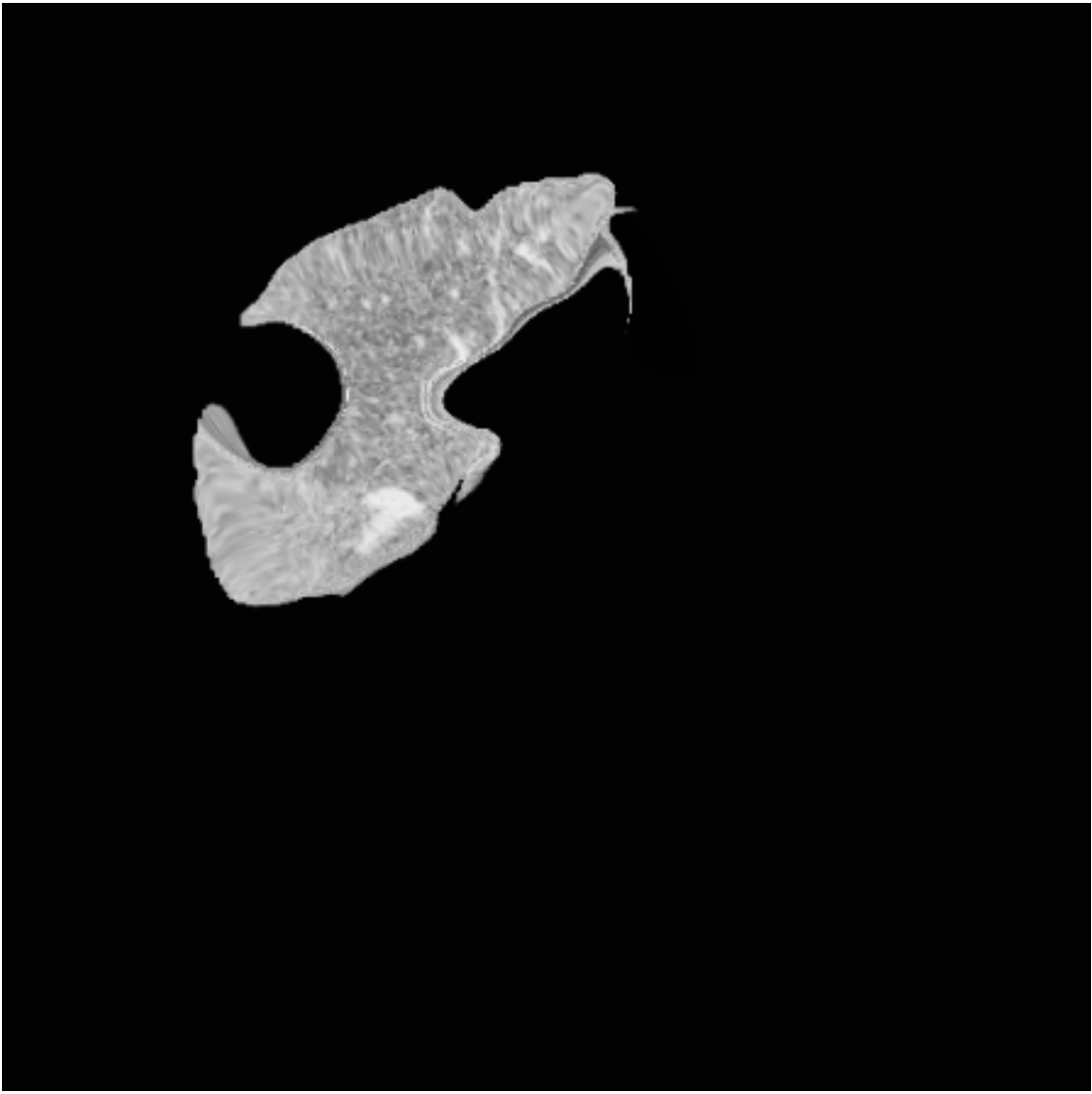}     &
        \includegraphics[width=0.17\textwidth]{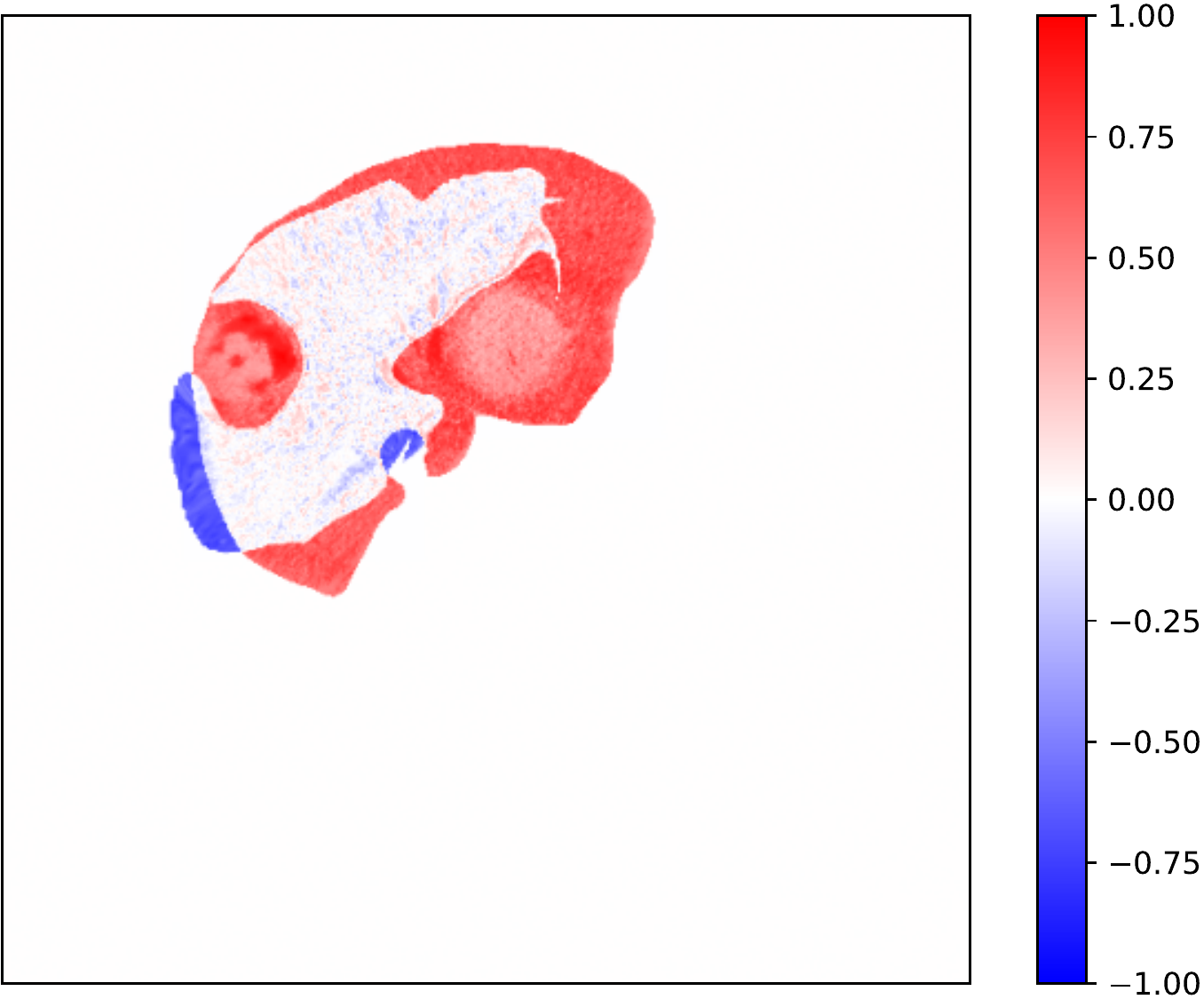} &
        \includegraphics[width=0.14\textwidth]{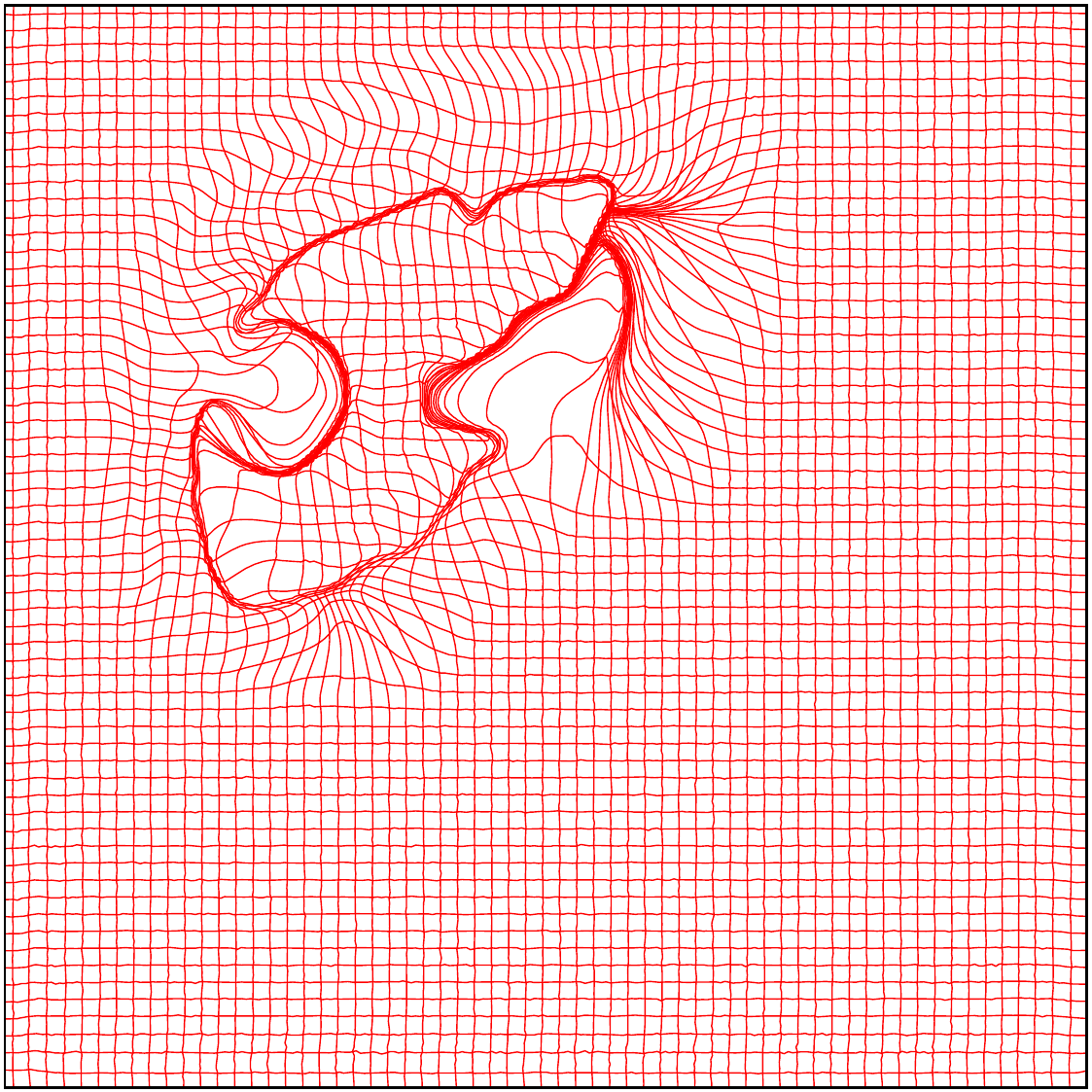}
        \\
        \rotatebox[origin=l]{90}{\scriptsize Ours} &
        \includegraphics[width=0.14\textwidth]{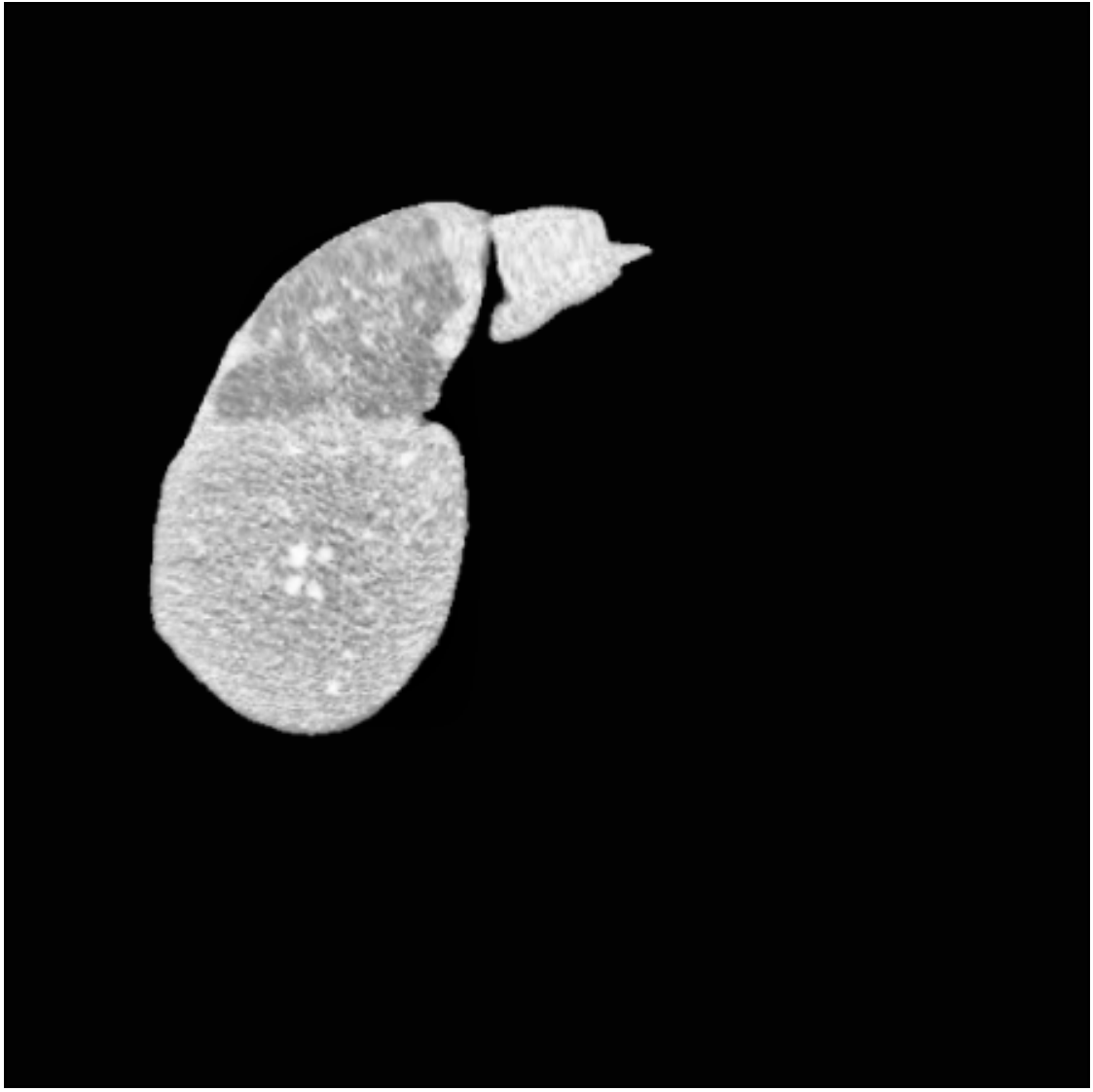}    &
        \includegraphics[width=0.17\textwidth]{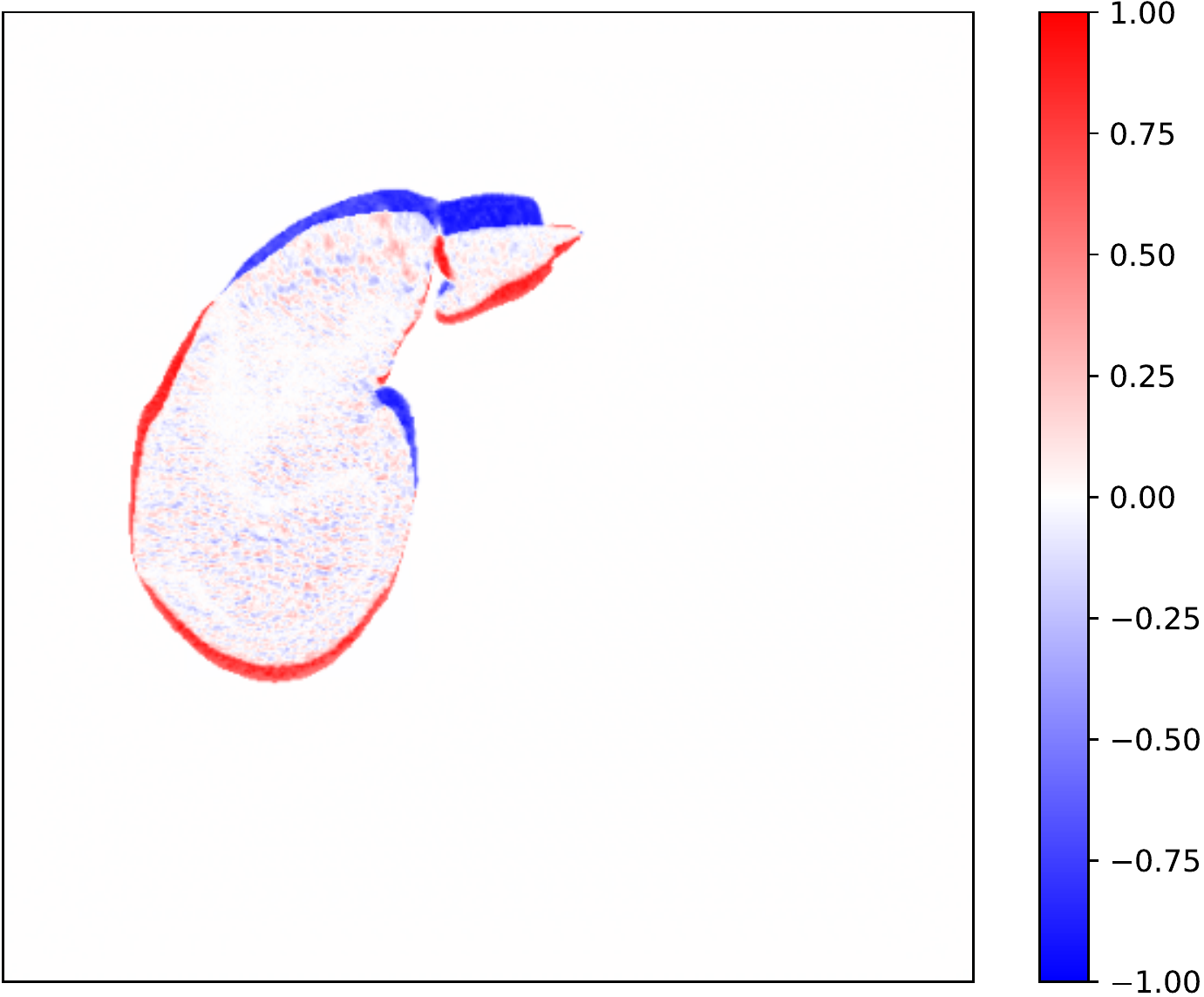} &
        \includegraphics[width=0.14\textwidth]{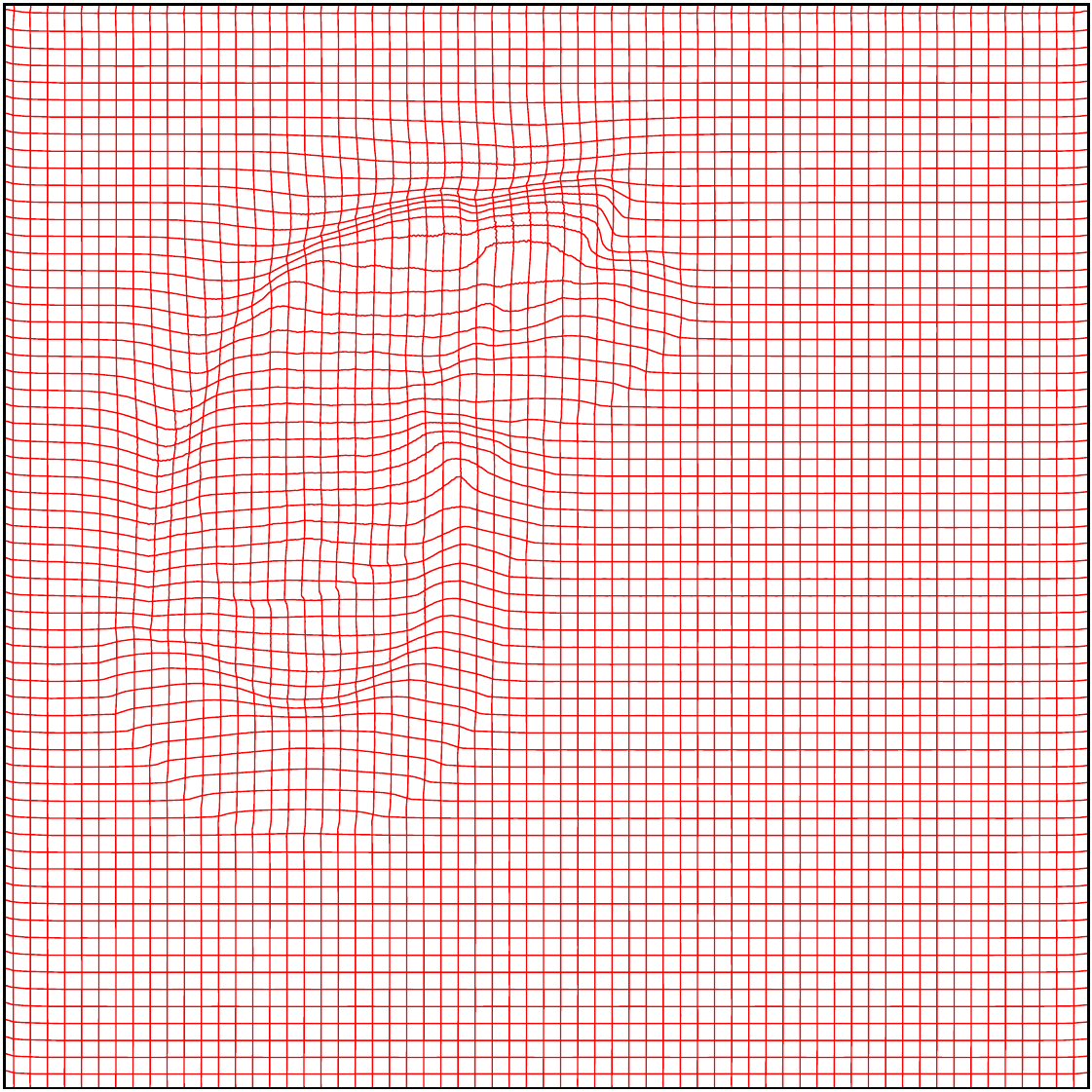} & \;
        \includegraphics[width=0.14\textwidth]{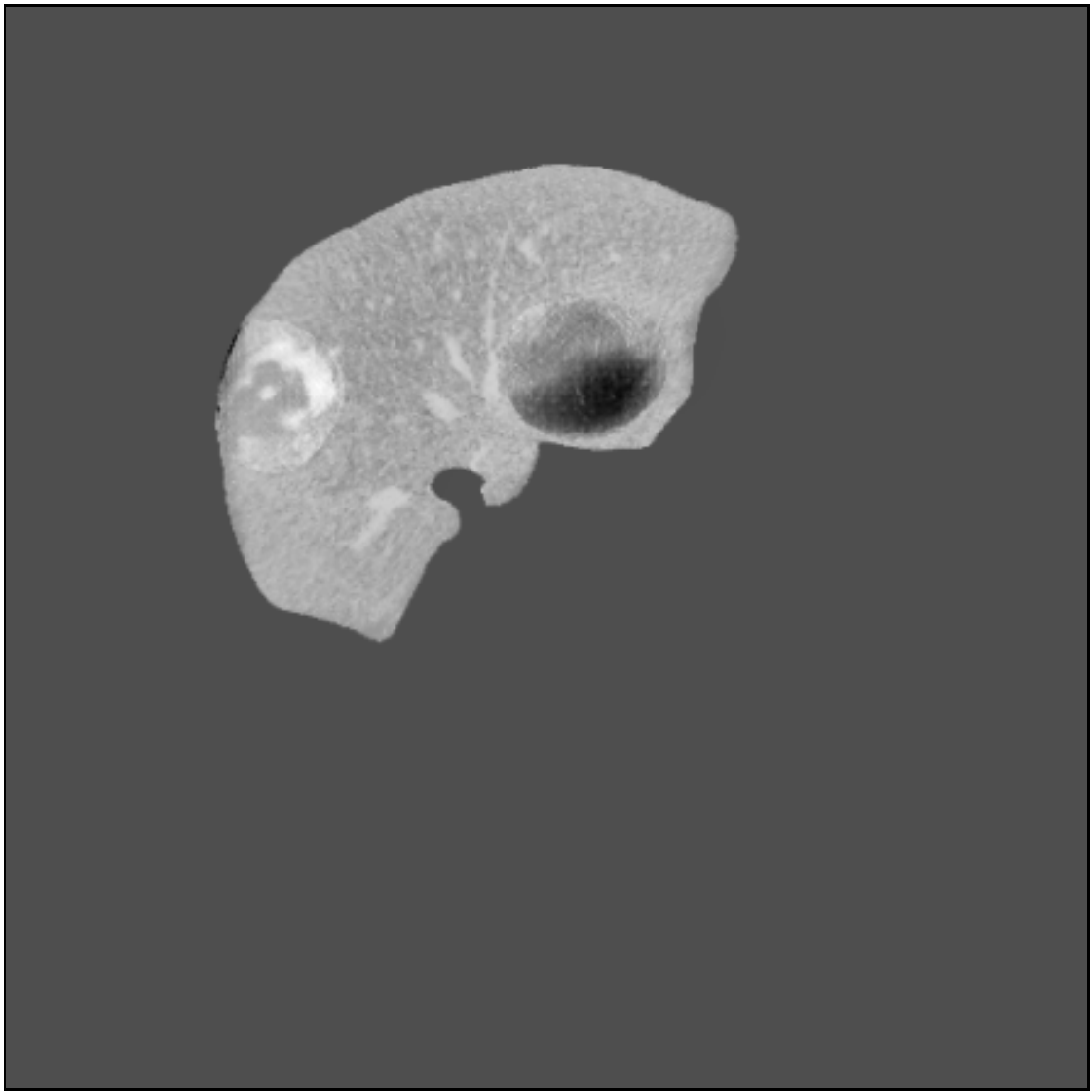}    &
        \includegraphics[width=0.17\textwidth]{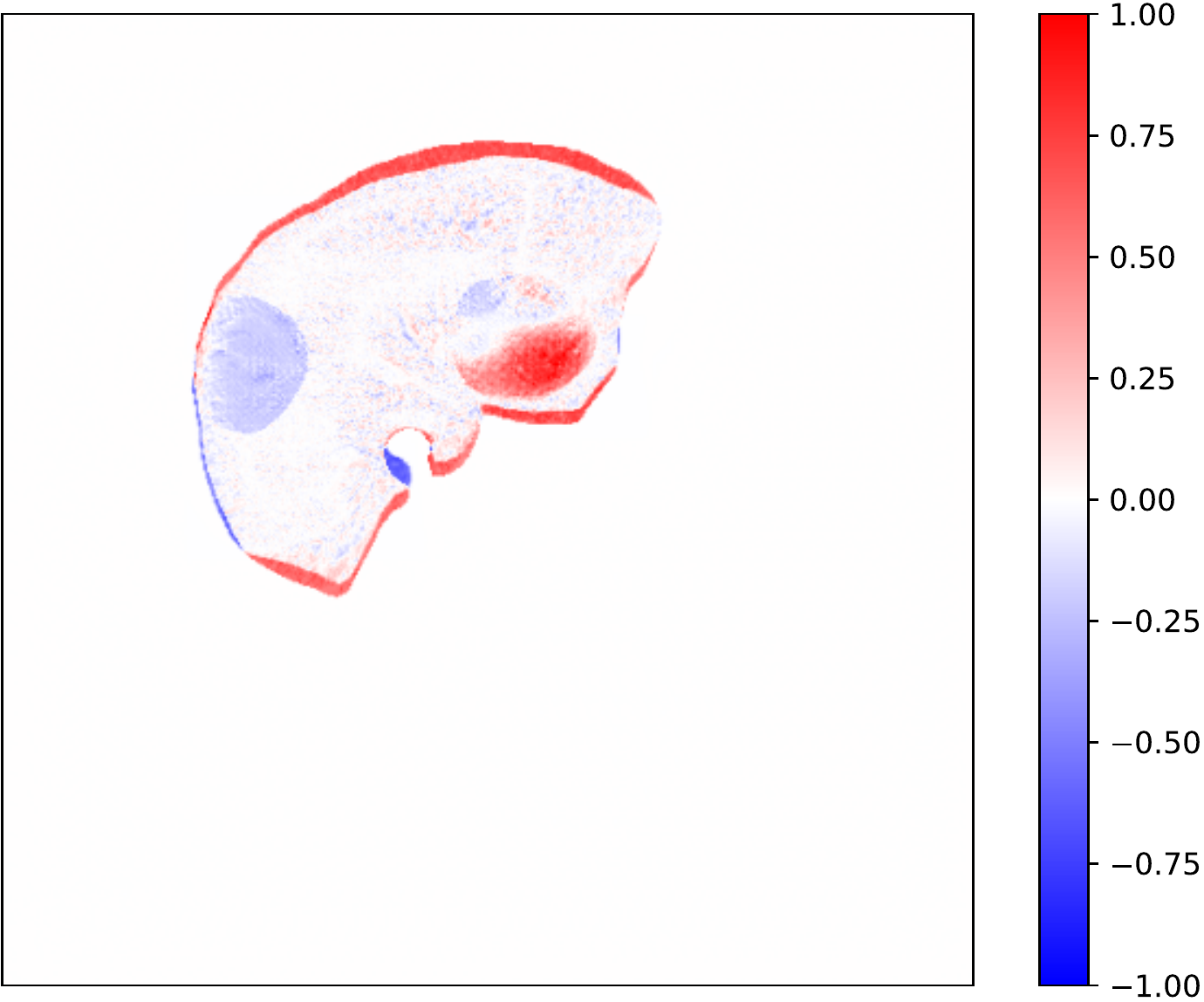} &
        \includegraphics[width=0.14\textwidth]{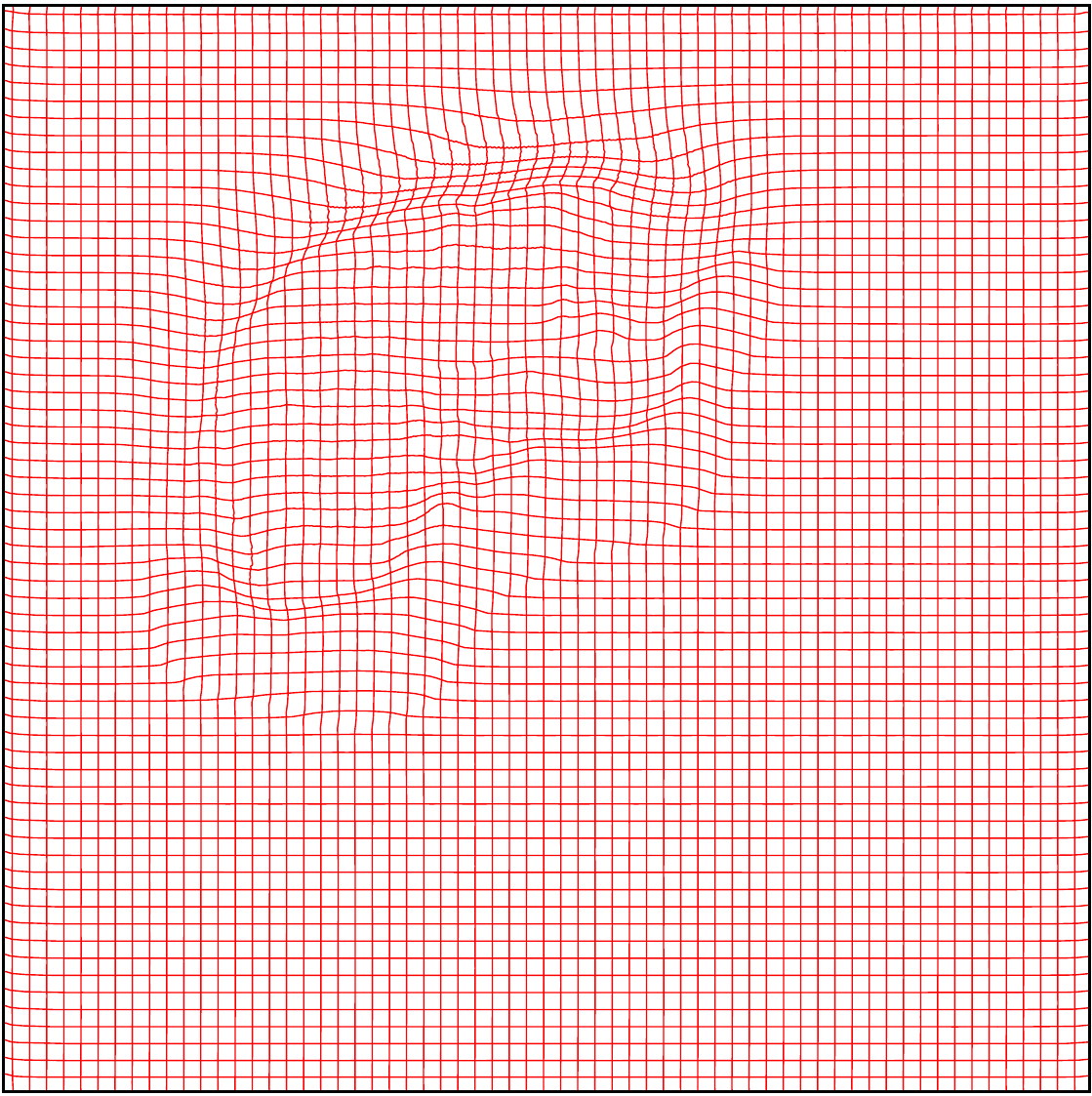}
    \end{tabular}
    \caption{Qualitative comparison between VM-CFM and our MetaRegNet on brain and liver datasets. Given an input pair on the top, the final output, its intensity difference to the target, and the spatial deformation are shown from left to right for each sample.}
    \label{fig:comparison_BraTS}
\end{figure}

\begin{table}[t]
  \caption{Comparison between VM-CFM and our method on brain and liver datasets.} 
    \label{tab:quant}
 \setlength{\tabcolsep}{3pt}
 \centering
  \begin{tabular}{ccccccc}
  
  \bfseries Data & \bfseries Method & \bfseries SSD ($e^{-1}$) & \bfseries SSD ($e^{-1}$) & \bfseries \#Foldings & \bfseries Time (ms) & \bfseries Dice \\
  & & SSD-total & SSD-healthy &  & (per img. pair)\\
 \hline  
 
Brain & VM-CFM & $0.13_{\pm0.003}$ & $0.1_{\pm 0.002}$ & $88.585_{\pm76.63}$ &21 & -\\
   & MetaRegNet & $\textbf{0.08}_{\pm0.002}$ & $\textbf{0.07}_{\pm0.002}$ & $\textbf{3.62}_{\pm6.34}$  & 23 & -\\
  \hline
   
   Liver & VM-CFM & $1.5_{\pm0.004}$ & $0.07_{\pm0.003}$  &  $23.16_{\pm34.56}$ & 22 & 0.88\\
 & MetaRegNet & $\textbf{0.40}_{\pm0.012}$ & $\textbf{0.04}_{\pm0.002}$ &  $\textbf{1.00}_{\pm4.90}$  & 24 &  \textbf{0.95} \\
 \hline
  \end{tabular}
\end{table}

\begin{figure}[t]
\centering
\setlength{\tabcolsep}{0.5pt}
    \begin{tabular}{llllll}
         $I_0$ & $I_1$ & $\mathcal{M}$ & $q(1) \otimes  \mathcal{M}$  & $\phi(1) \circ I_0$ & Final  \\
           \includegraphics[width=0.15\textwidth]{figures/output4/source.pdf}  &
            \includegraphics[width=0.15\textwidth]{figures/output4/target.pdf}  & \includegraphics[width=0.15\textwidth]{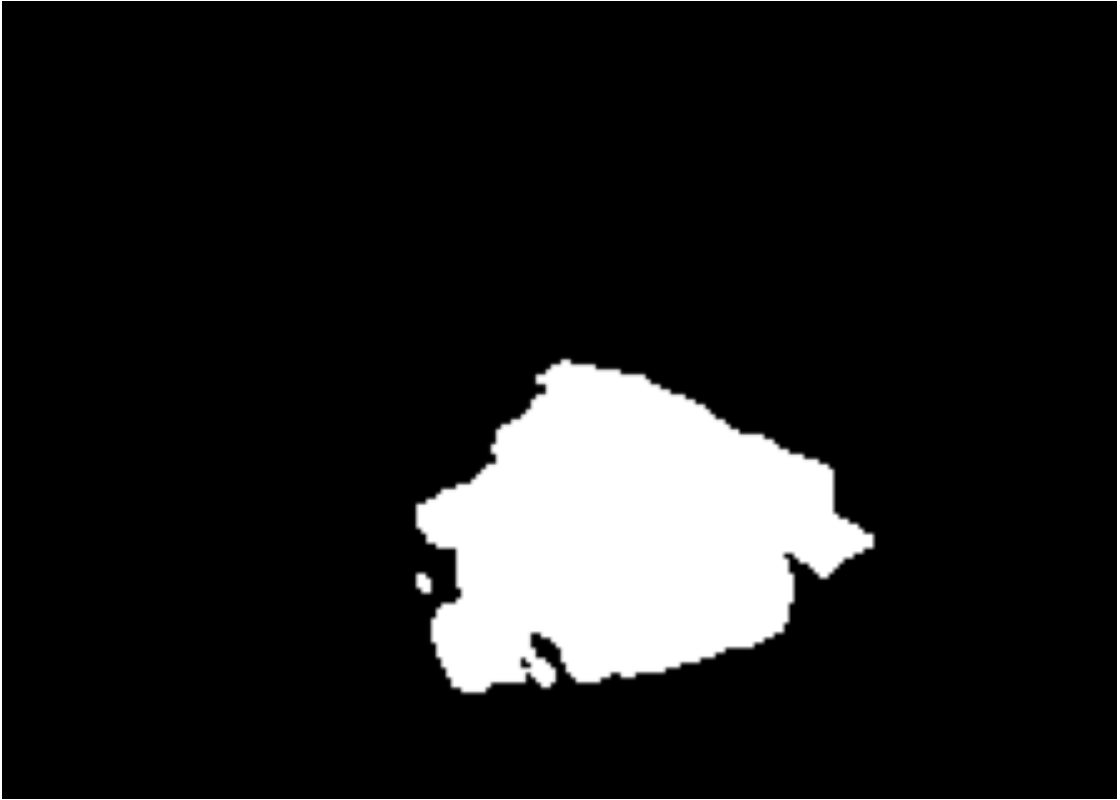}    &
            \includegraphics[width=0.15\textwidth]{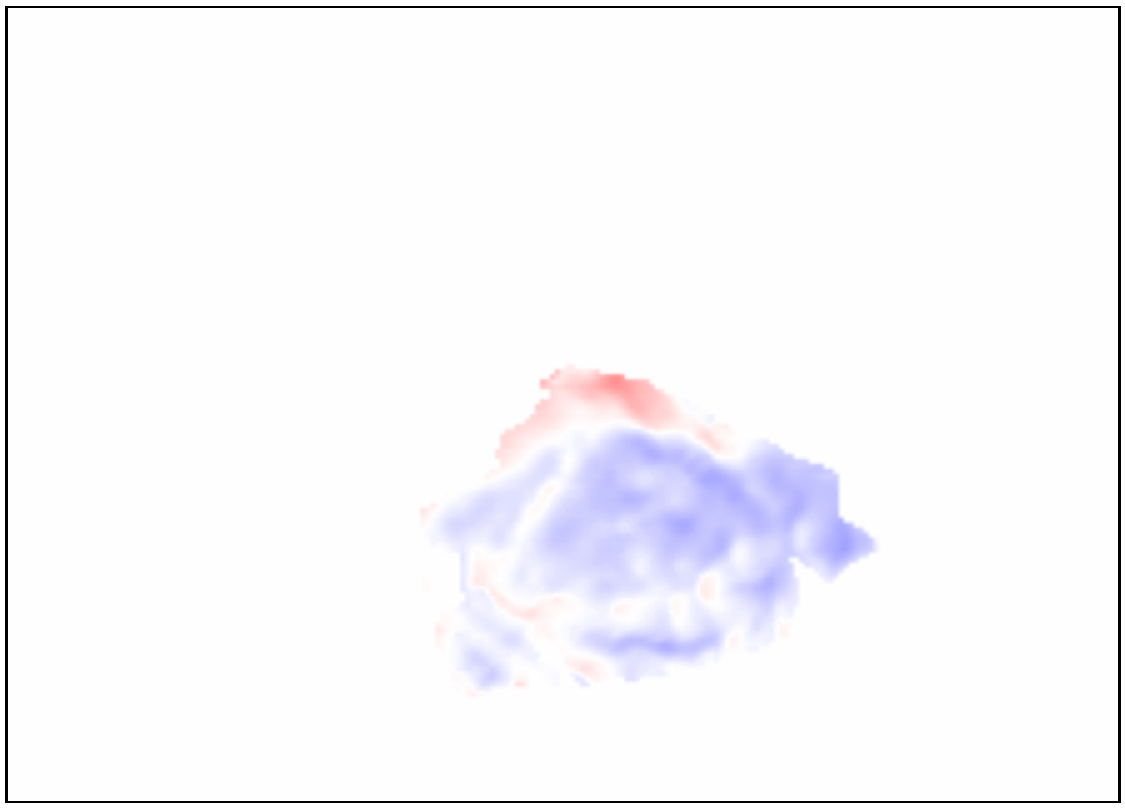} &
            \includegraphics[width=0.15\textwidth]{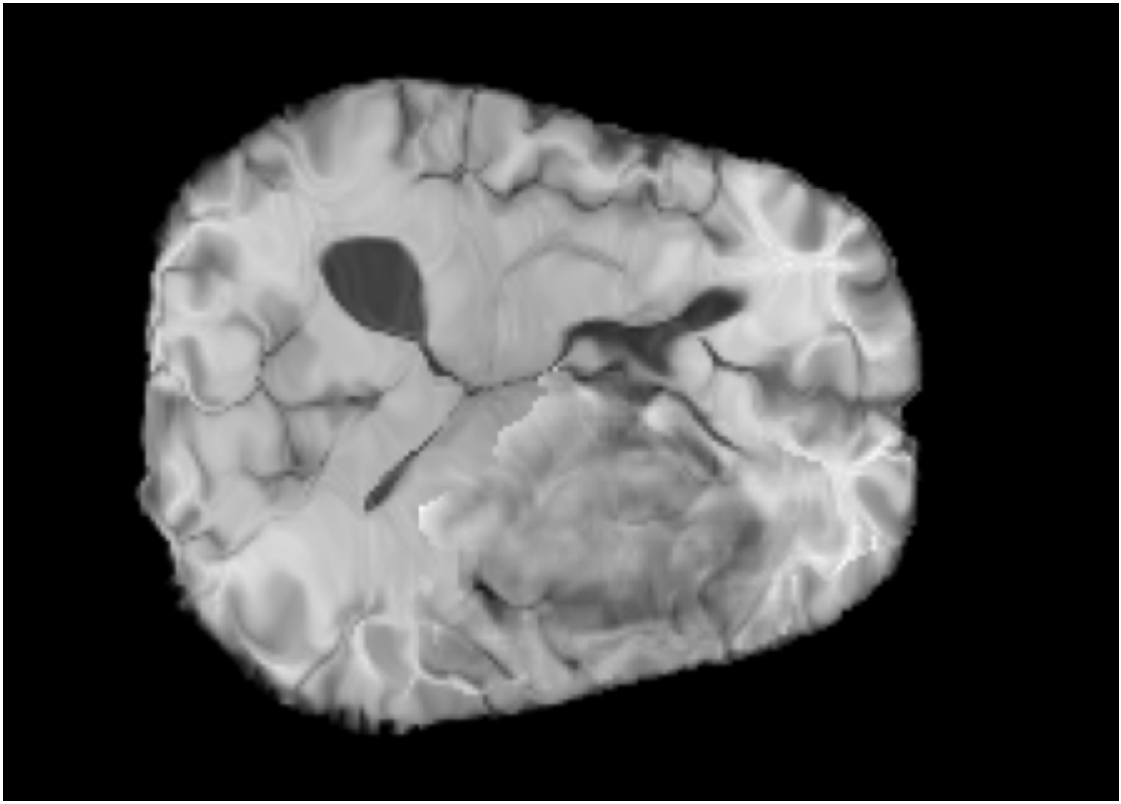} &
             \includegraphics[width=0.15\textwidth]{figures/output4/nsvf.pdf} \\ 
             
          \includegraphics[width=0.15\textwidth]{liver/output1/source.pdf}  & 
        \includegraphics[width=0.15\textwidth]{liver/output1/target.pdf}  & \includegraphics[width=0.15\textwidth]{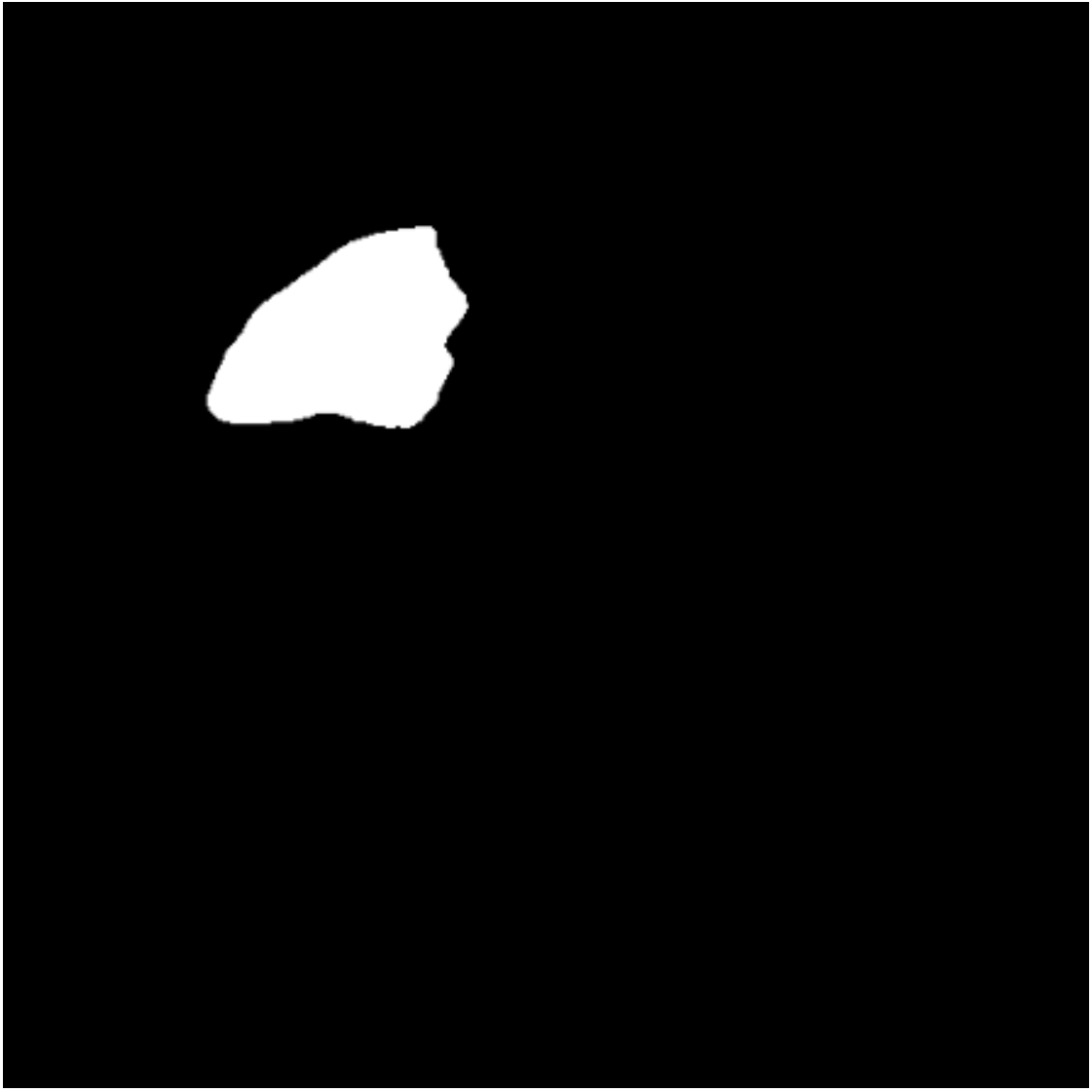}     &
        \includegraphics[width=0.185\textwidth]{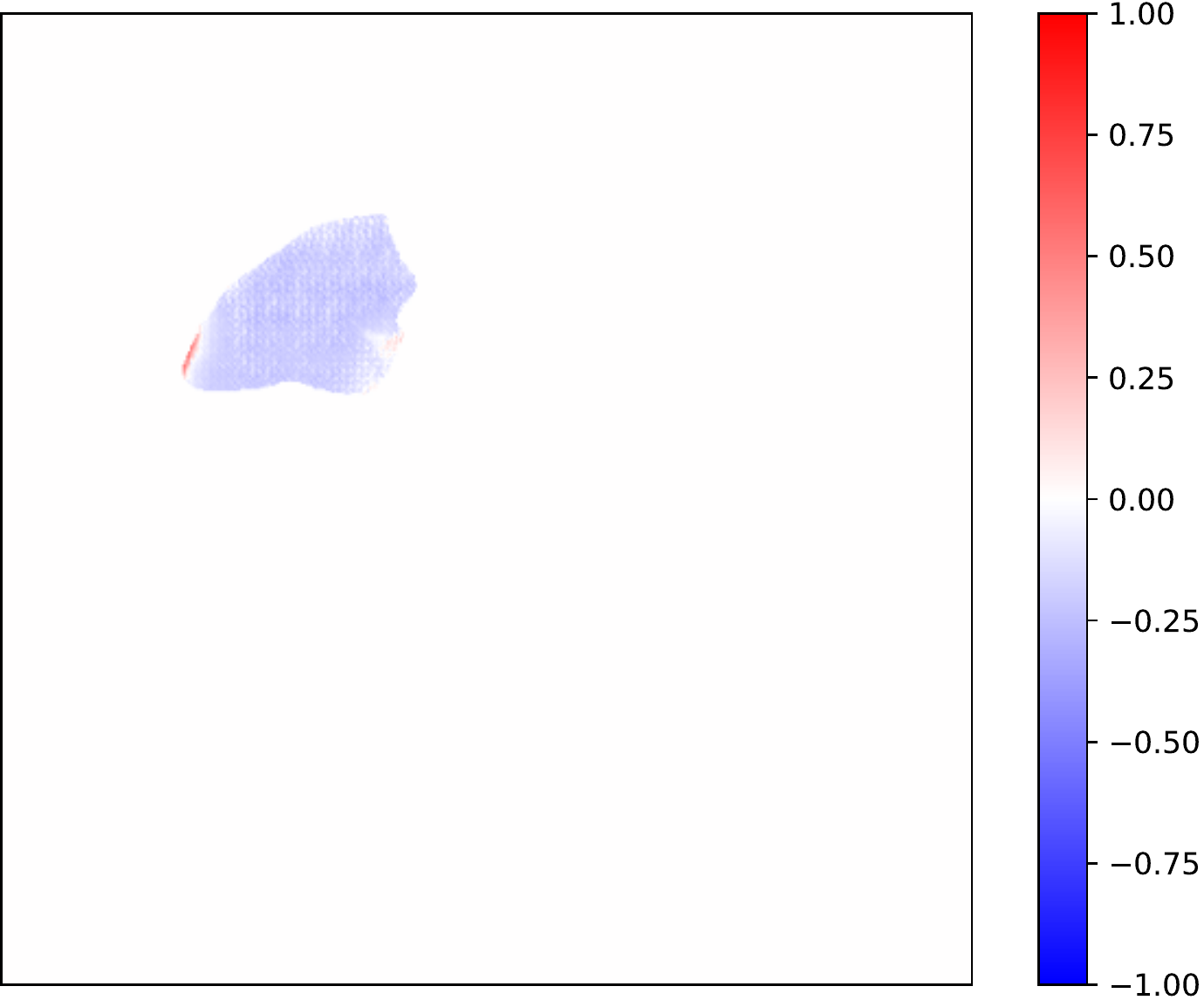} &
          \includegraphics[width=0.15\textwidth]{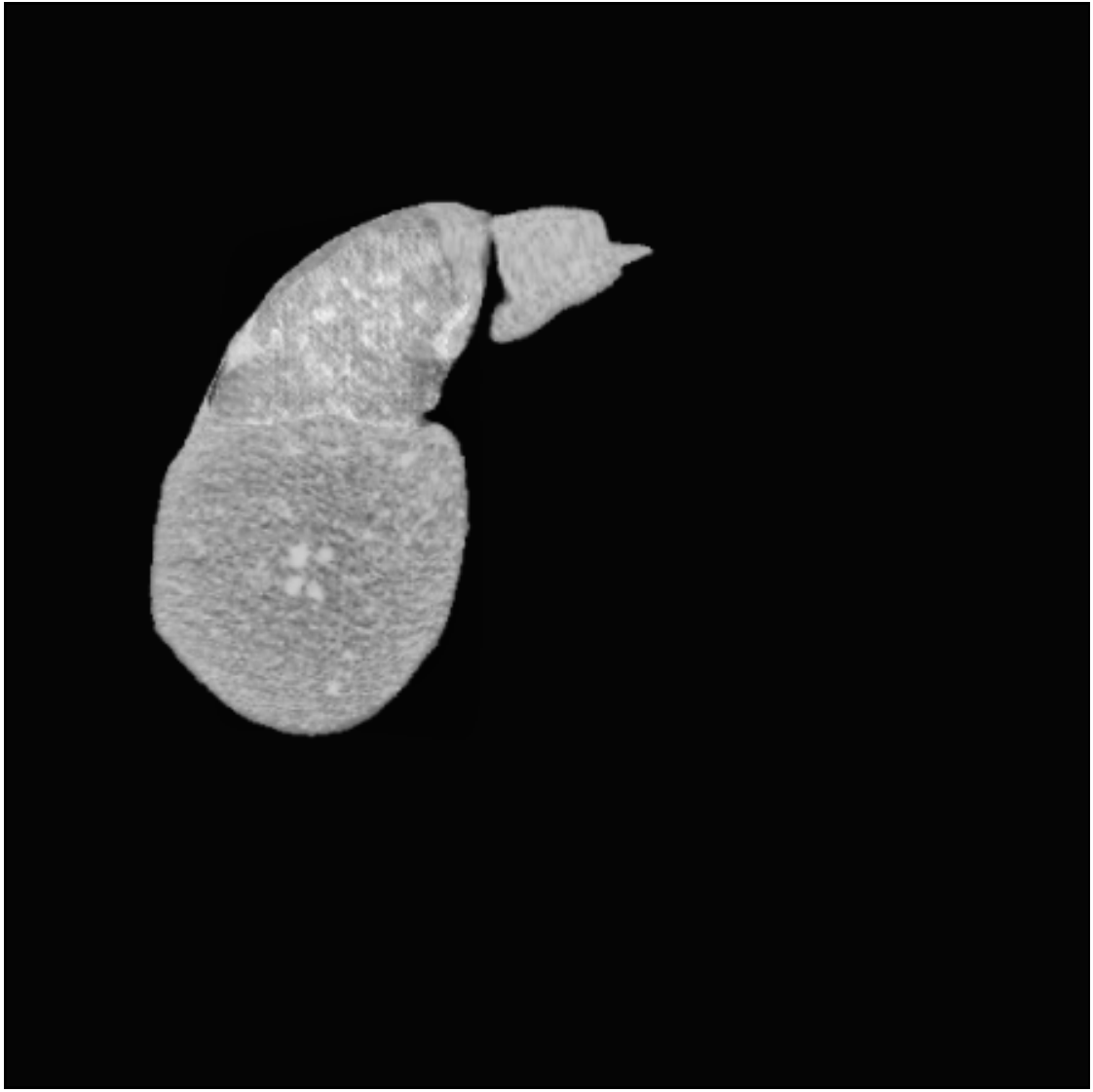} &
        \includegraphics[width=0.15\textwidth]{liver/output1/nsvf.pdf} \\
            
    \end{tabular}
    \caption{Image appearance separation learned by MetaRegNet. Left to right: clean source image $I_{0}$, pathological target image $I_{1}$, tumor mask $\mathcal{M}$, learned intensity changes within the tumor region $q(1) \otimes \mathcal{M}$, the deformed source image $\phi(1) \circ I_0$, and the final output.}
    \label{fig:appearance}
\end{figure}

\noindent
\textbf{Experimental Results.}
Table~\ref{tab:quant} reports our quantitative results. Compared to VM-CFM, our model provides a better matching, not only in the healthy region but also in the whole image, for both brain and liver datasets. Fig.~\ref{fig:comparison_BraTS} presents the visual improvement, showing better matching in both tumor and healthy regions, for images with either large or small tumors. Our deformations are much smoother, as demonstrated by the much fewer foldings reported in Table~\ref{tab:quant} and smoother maps within tumors and their surrounding regions in Fig.~\ref{fig:comparison_BraTS}. 
Unlike VM-CFM, we have spatial deformations going under the tumor regions, and the rest appearance changes are contributed by the intensity variance, indicating the appearing tumors from a healthy source to a pathological target, which is also observed in Fig.~\ref{fig:appearance}.
To further evaluate the effectiveness of MetaRegNet, we use deformation maps to transfer the segmentation mask of the source image to match the target one. Since only liver masks are available, we apply our method on liver segmentation and obtain a mean Dice score of 0.95, compared to 0.88 produced by VM-CFM. And we only takes 2ms more for registering one pair.

\section{Conclusion and Future Work}


In this paper, we propose a metamorphic image registration framework, MetaRegNet, which utilizes LC-ResNet blocks as flow integrator and allows for joint estimation of spatial deformation and intensity variation between a healthy source image and a pathological target image. MetaRegNet is successfully applied on two real datasets with big tumors or multiple ones in the same scan, by producing more realistic matching results and smoother deformations compared to exist methods. 
Limited by few available 3D medical datasets with both normal and tumor images, we currently work on 2D images. However, the model is developed for 3D metamorphic image registration and we will perform experiments on $3D$ medical scans in future work, as well as other applications with considering more anatomies and image modalities. Another possible extension of our work is the adaption to a model that registers any pairs of healthy and pathological images, including the registration between two pathological image scans. 


%
%
%


\bibliographystyle{splncs04}
\bibliography{refs}

\begin{thebibliography}{10}
\providecommand{\url}[1]{\texttt{#1}}
\providecommand{\urlprefix}{URL }
\providecommand{\doi}[1]{https://doi.org/#1}

\bibitem{3dircadb}
3dircadb-01. \url{https://www.ircad.fr/research/3dircadb/}

\bibitem{almotairi2020liver}
Almotairi, S., Kareem, G., Aouf, M., Almutairi, B., Salem, M.A.M.: Liver tumor
  segmentation in ct scans using modified segnet. Sensors  \textbf{20}(5),
  ~1516 (2020)

\bibitem{arsigny2006log}
Arsigny, V., Commowick, O., Pennec, X., Ayache, N.: A log-euclidean framework
  for statistics on diffeomorphisms. In: International Conference on Medical
  Image Computing and Computer-Assisted Intervention. pp. 924--931. Springer
  (2006)

\bibitem{avants2011reproducible}
Avants, B.B., Tustison, N.J., Song, G., Cook, P.A., Klein, A., Gee, J.C.: A
  reproducible evaluation of ants similarity metric performance in brain image
  registration. Neuroimage  \textbf{54}(3),  2033--2044 (2011)

\bibitem{baid2021rsna}
Baid, U., Ghodasara, S., Mohan, S., Bilello, M., Calabrese, E., Colak, E.,
  Farahani, K., Kalpathy-Cramer, J., Kitamura, F.C., Pati, S., et~al.: The
  rsna-asnr-miccai brats 2021 benchmark on brain tumor segmentation and
  radiogenomic classification. arXiv preprint arXiv:2107.02314  (2021)

\bibitem{beg2005computing}
Beg, M.F., Miller, M.I., Trouv{\'e}, A., Younes, L.: Computing large
  deformation metric mappings via geodesic flows of diffeomorphisms.
  International journal of computer vision  \textbf{61}(2),  139--157 (2005)

\bibitem{bone2020learning}
B{\^o}ne, A., Vernhet, P., Colliot, O., Durrleman, S.: Learning joint shape and
  appearance representations with metamorphic auto-encoders. In: International
  Conference on Medical Image Computing and Computer-Assisted Intervention. pp.
  202--211. Springer (2020)

\bibitem{brett2001spatial}
Brett, M., Leff, A.P., Rorden, C., Ashburner, J.: Spatial normalization of
  brain images with focal lesions using cost function masking. Neuroimage
  \textbf{14}(2),  486--500 (2001)

\bibitem{cao2005large}
Cao, Y., Miller, M.I., Winslow, R.L., Younes, L.: Large deformation
  diffeomorphic metric mapping of vector fields. TMI  \textbf{24}(9),
  1216--1230 (2005)

\bibitem{chitphakdithai2010non}
Chitphakdithai, N., Duncan, J.S.: Non-rigid registration with missing
  correspondences in preoperative and postresection brain images. In: MICCAI.
  pp. 367--374. Springer (2010)

\bibitem{clatz2005robust}
Clatz, O., Delingette, H., Talos, I.F., Golby, A.J., Kikinis, R., Jolesz, F.A.,
  Ayache, N., Warfield, S.K.: Robust nonrigid registration to capture brain
  shift from intraoperative mri. IEEE transactions on medical imaging
  \textbf{24}(11),  1417--1427 (2005)

\bibitem{dalca2018unsupervised}
Dalca, A.V., Balakrishnan, G., Guttag, J., Sabuncu, M.R.: Unsupervised learning
  for fast probabilistic diffeomorphic registration. In: MICCAI. pp. 729--738.
  Springer (2018)

\bibitem{franccois2021metamorphic}
Fran{\c{c}}ois, A., Gori, P., Glaun{\`e}s, J.: Metamorphic image registration
  using a semi-lagrangian scheme. In: International Conference on Geometric
  Science of Information. pp. 781--788. Springer (2021)

\bibitem{franccois2022weighted}
Fran{\c{c}}ois, A., Maillard, M., Oppenheim, C., Pallud, J., Gori, P.,
  Glaun{\`e}s, J.A.: Weighted metamorphosis for registration of images with
  different topology. In: 10th International Workshop on Biomedical Image
  Registration (2022)

\bibitem{gooya2011joint}
Gooya, A., Pohl, K.M., Bilello, M., Biros, G., Davatzikos, C.: Joint
  segmentation and deformable registration of brain scans guided by a tumor
  growth model. In: MICCAI. pp. 532--540. Springer (2011)

\bibitem{gooya2012glistr}
Gooya, A., Pohl, K.M., Bilello, M., Cirillo, L., Biros, G., Melhem, E.R.,
  Davatzikos, C.: Glistr: glioma image segmentation and registration. IEEE
  transactions on medical imaging  \textbf{31}(10),  1941--1954 (2012)

\bibitem{han2018brain}
Han, X., Kwitt, R., Aylward, S., Bakas, S., Menze, B., Asturias, A., Vespa, P.,
  Van~Horn, J., Niethammer, M.: Brain extraction from normal and pathological
  images: A joint pca/image-reconstruction approach. NeuroImage  \textbf{176},
  431--445 (2018)

\bibitem{han2020deep}
Han, X., Shen, Z., Xu, Z., Bakas, S., Akbari, H., Bilello, M., Davatzikos, C.,
  Niethammer, M.: A deep network for joint registration and reconstruction of
  images with pathologies. In: International Workshop on Machine Learning in
  Medical Imaging. pp. 342--352. Springer (2020)

\bibitem{hong2012metamorphic}
Hong, Y., Joshi, S., Sanchez, M., Styner, M., Niethammer, M.: Metamorphic
  geodesic regression. In: International Conference on Medical Image Computing
  and Computer-Assisted Intervention. pp. 197--205. Springer (2012)

\bibitem{joshi2021diffeomorphic}
Joshi, A., Hong, Y.: Diffeomorphic image registration using lipschitz
  continuous residual networks. In: Medical Imaging with Deep Learning (2021)

\bibitem{kuzilek2017open}
Kuzilek, J., Hlosta, M., Zdrahal, Z.: Open university learning analytics
  dataset. Scientific data  \textbf{4}(1), ~1--8 (2017)

\bibitem{liu2014low}
Liu, X., Niethammer, M., Kwitt, R., McCormick, M., Aylward, S.: Low-rank to the
  rescue--atlas-based analyses in the presence of pathologies. In: MICCAI. pp.
  97--104. Springer (2014)

\bibitem{maillard2022deep}
Maillard, M., Fran{\c{c}}ois, A., Glaun{\`e}s, J., Bloch, I., Gori, P.: A deep
  residual learning implementation of metamorphosis. In: 2022 IEEE 19th
  International Symposium on Biomedical Imaging (ISBI). pp.~1--4. IEEE (2022)

\bibitem{menze2014multimodal}
Menze, B.H., Jakab, A., Bauer, S., Kalpathy-Cramer, J., Farahani, K., Kirby,
  J., Burren, Y., Porz, N., Slotboom, J., Wiest, R., et~al.: The multimodal
  brain tumor image segmentation benchmark (brats). IEEE transactions on
  medical imaging  \textbf{34}(10),  1993--2024 (2014)

\bibitem{menze2009menze}
Menze, J., Masuch, H., Bachert, P., et~al.: Menze bh. Kelm BM, Masuch R.,
  Himmelreich U., Bachert P., Petrich W., et al., A comparison of random forest
  and its gini importance with standard chemometric methods for the feature
  selection and classification of spectral data, BMC Bioinformatics
  \textbf{10}(1) (2009)

\bibitem{mok2020fast}
Mok, T.C., Chung, A.: Fast symmetric diffeomorphic image registration with
  convolutional neural networks. In: Proceedings of the IEEE/CVF conference on
  computer vision and pattern recognition. pp. 4644--4653 (2020)

\bibitem{mok2022unsupervised}
Mok, T.C., Chung, A.: Unsupervised deformable image registration with absent
  correspondences in pre-operative and post-recurrence brain tumor mri scans.
  arXiv preprint arXiv:2206.03900  (2022)

\bibitem{niethammer2011geometric}
Niethammer, M., Hart, G.L., Pace, D.F., Vespa, P.M., Irimia, A., Horn, J.D.V.,
  Aylward, S.R.: Geometric metamorphosis. In: MICCAI. pp. 639--646. Springer
  (2011)

\bibitem{ronneberger2015u}
Ronneberger, O., Fischer, P., Brox, T.: U-net: Convolutional networks for
  biomedical image segmentation. In: International Conference on Medical image
  computing and computer-assisted intervention. pp. 234--241. Springer (2015)

\bibitem{sdika2009nonrigid}
Sdika, M., Pelletier, D.: Nonrigid registration of multiple sclerosis brain
  images using lesion inpainting for morphometry or lesion mapping. Tech. rep.,
  Wiley Online Library (2009)

\bibitem{shu2018deforming}
Shu, Z., Sahasrabudhe, M., Guler, R.A., Samaras, D., Paragios, N., Kokkinos,
  I.: Deforming autoencoders: Unsupervised disentangling of shape and
  appearance. In: Proceedings of ECCV. pp. 650--665 (2018)

\bibitem{sotiras2013deformable}
Sotiras, A., Davatzikos, C., Paragios, N.: Deformable medical image
  registration: A survey. IEEE transactions on medical imaging  \textbf{32}(7),
   1153--1190 (2013)

\bibitem{trouve2005metamorphoses}
Trouv{\'e}, A., Younes, L.: Metamorphoses through lie group action. Foundations
  of computational mathematics  \textbf{5}(2),  173--198 (2005)

\bibitem{yang2017quicksilver}
Yang, X., Kwitt, R., Styner, M., Niethammer, M.: Quicksilver: Fast predictive
  image registration--a deep learning approach. NeuroImage  \textbf{158},
  378--396 (2017)

\end{thebibliography}

\end{document}